\documentclass[traditabstract]{aa}
\usepackage{graphicx}
\usepackage{txfonts}
\usepackage{amssymb}
\usepackage[below]{placeins}
\usepackage{natbib}
\usepackage{longtable,lscape}
\bibpunct{(}{)}{;}{a}{}{,}
\usepackage{grffile}
\usepackage{epstopdf}
\graphicspath{{}}
\usepackage[table]{xcolor}
\definecolor{tableShade}{HTML}{E4E4E4}
\usepackage{float}
\usepackage{perpage}
\MakeSorted{figure}
\MakeSorted{table}

\begin{document}

\title{Evolution of the galaxy luminosity function in progenitors of fossil groups}
\titlerunning{A significant evidence for the origin of fossil groups}
\authorrunning{Gozaliasl et al. }
 \author{
 G. Gozaliasl\inst{1,2}
 \and
 H. G. Khosroshahi\inst{3}
 \and
  A. A. Dariush\inst{4}
 \and
  A. Finoguenov\inst{1}
\and    
 D.M.Z. Jassur \inst{2}
 \and  
 A. Molaeinajad \inst{3}  }
 
\institute{
Department of Physics, University of Helsinki, P. O. Box 64, FI-00014, Helsinki, Finland\\
\email{ghassem.gozaliasl@helsinki.fi}
\and
Department of Theoretical Physics and Astrophysics, University of Tabriz, P.O. Box 51664, Tabriz, Iran
\and
School of Astronomy, Institute for Research in Fundamental Sciences (IPM), Tehran, Iran
\and
Institute of Astronomy, University of Cambridge, Madingley Road, Cambridge CB3 0HA, UK
}

\date{Received ....; accepted ....}

\abstract {Using the semi-analytic models based on the Millennium
  simulation, we trace back the evolution of the luminosity function of
  galaxies residing in progenitors of groups classified by the
  magnitude gap at redshift zero. We determine the luminosity
  function of galaxies within $ 0.25R_{200}, 0.5R_{200} $, and $
  R_{200} $ for galaxy groups/clusters. The bright end of the galaxy luminosity 
  function of fossil groups shows a significant evolution with redshift, with changes in $M^*$ by $\sim$ 1-2 
  mag between $z\sim0.5$ and $z=0$ (for the central $0.5R_{200}$), suggesting
  that the formation of the most luminous galaxy in a fossil group has 
  had a significant impact on the $M^{*}$ galaxies e.g. it is formed as a result of  
  multiple mergers of $ M^{*} $ galaxies within the last 5 Gyr. In
  contrast, the slope of the faint end, $\alpha$, of the luminosity
  function shows no considerable redshift evolution and the number of
  dwarf galaxies in the fossil groups exhibits no evolution, unlike in
  non-fossil groups where it grows by $\sim25-42$\% towards low
  redshifts.  In agreement with previous studies, we also show that
  fossil groups accumulate most of their halo mass earlier than
  non-fossil groups.
  Selecting the fossils at a redshift of 1 and tracing them to a redshift
  0, we show that 80\% of the fossil groups ($10^{13} M_{\odot} h^{-1}<
  M_{200}<10^{14} M_{\odot} h^{-1}$) will lose their large magnitude 
  gaps.  However, about 40\% of fossil clusters ($M_{200}>10^{14}$
  $M_{\odot}$ $h^{-1} $) will retain their large gaps.

}

 

 \maketitle
\section{Introduction}

\label{sec:intro}
The first fossil galaxy group, $ RX J1340.6+4018 $, was discovered by
\cite{ponman94}.  Since then, a number of studies have revealed that
these systems exhibit interesting properties compared to similar
normal groups and clusters
\cite[e.g.][]{vikhlinin99,jones03,kpj06,kmpj06,khosroshahi07,dariush07,Miller12,Proctor11,Eigenthaler13,kjp04,labarbera09}.

According to the first formal definition presented by \cite{jones03},
a fossil group of galaxies is identified as an extended X-ray 
source with L$_{X, bol} \geqslant$2.13 $ \times$10$^{42} $ h$_{50}^{-2}$ergs$^{-1}$ that includes a luminous elliptical galaxy with at least
two magnitude difference in the R-band from the second brightest galaxy
located within half the virial radius.
 
\cite{jones03} reported 8\% to 20 \% of all galaxy groups  have the same X-ray luminosity 
 as fossil groups. The redshift
evolution of the fossil fraction has been studied by
\citet{Gozaliasl14}, finding that fossil groups constitute $22\pm6$\%
and $13\pm7$\% of massive galaxy groups ($M_{200}\sim 10^{13.5} M_{\sun} h^{-1}$) at
$z\leq0.6$ and $0.6<z<1.2$, respectively. In the semi-analytic halo catalogue, 
based on the Millennium simulation, $\sim$7.2$ \pm $0.2 per cent of
massive groups are fossils \citep{dariush07}. 

Today, there is no consensus about the origin of fossil groups
\citep{Mulchaey99,cypriano06,ponman94,Aguerri11,Proctor11,Girardi14,Zarattini14}.
\cite{Mulchaey99} portrayed a fossil group as a ``failed group'' in
which the majority of the baryonic materials were used in building up
the giant elliptical galaxy, rather than in several massive $L ^{*} $
galaxies. In contrast, following \cite{barnes89}, number of studies
have suggested that the central elliptical galaxy in a fossil group
forms because of  multiple mergers of massive galaxies in groups in a few
tenths of a Hubble time. For this to be a viable formation scenario,
we expect fossil groups to present a luminosity function (LF) with a
deficit of $L ^{*} $ galaxies. The key lies in the timescale for
the merging of dwarf galaxies and cooling group X-ray halo 
\citep{ponman93,ponman94}, which is much longer than the timescale for
the merging of more luminous galaxies. The final product of these
mergers are present day isolated ellipticals, which are immersed in
group size X-ray halos and surrounded by substantially fainter
galaxies \citep[e.g.][]{ponman94,donghia05}.
 \begin{figure*}
        \begin{center}  
          \leavevmode
          \includegraphics[width=14cm]{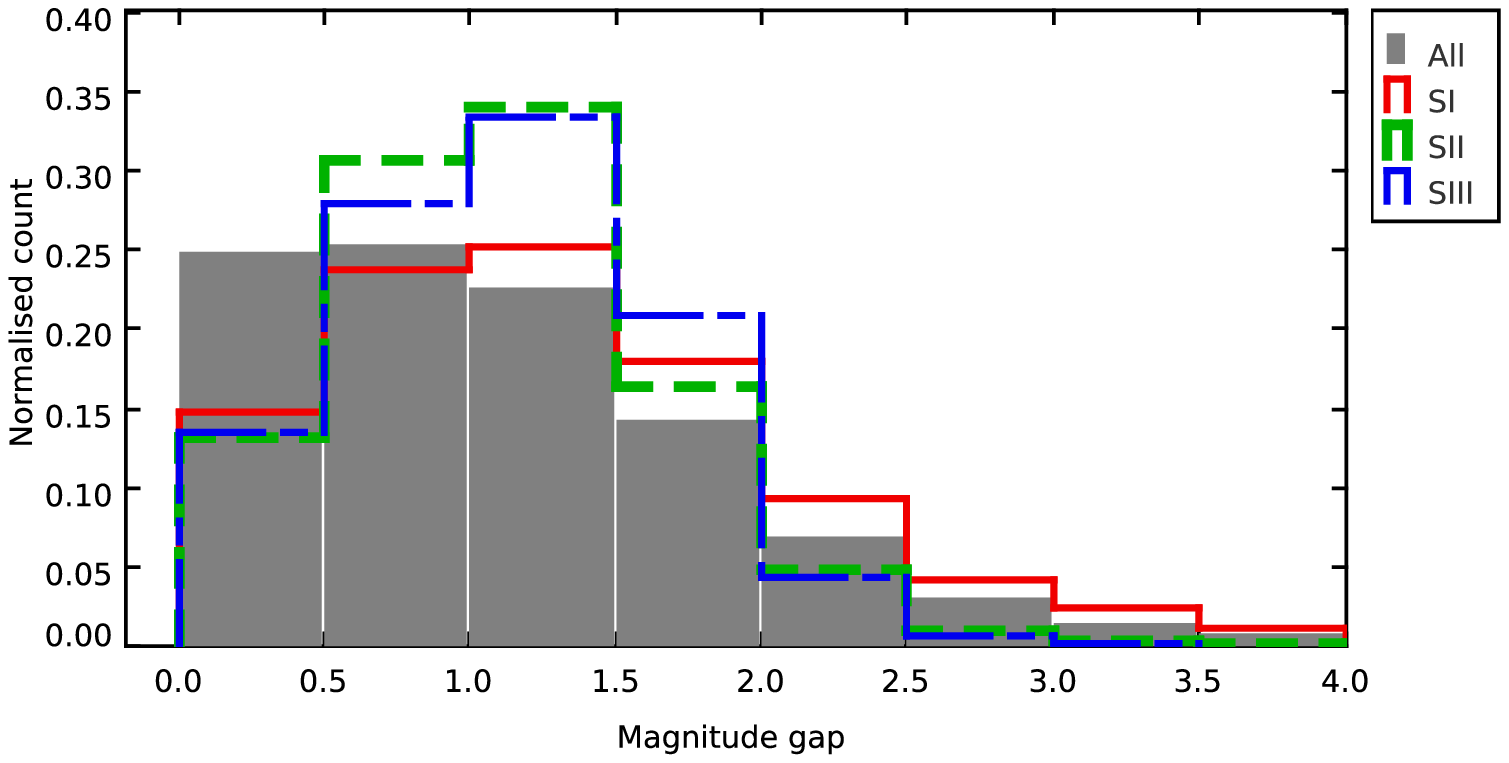}
          \includegraphics[width=14cm]{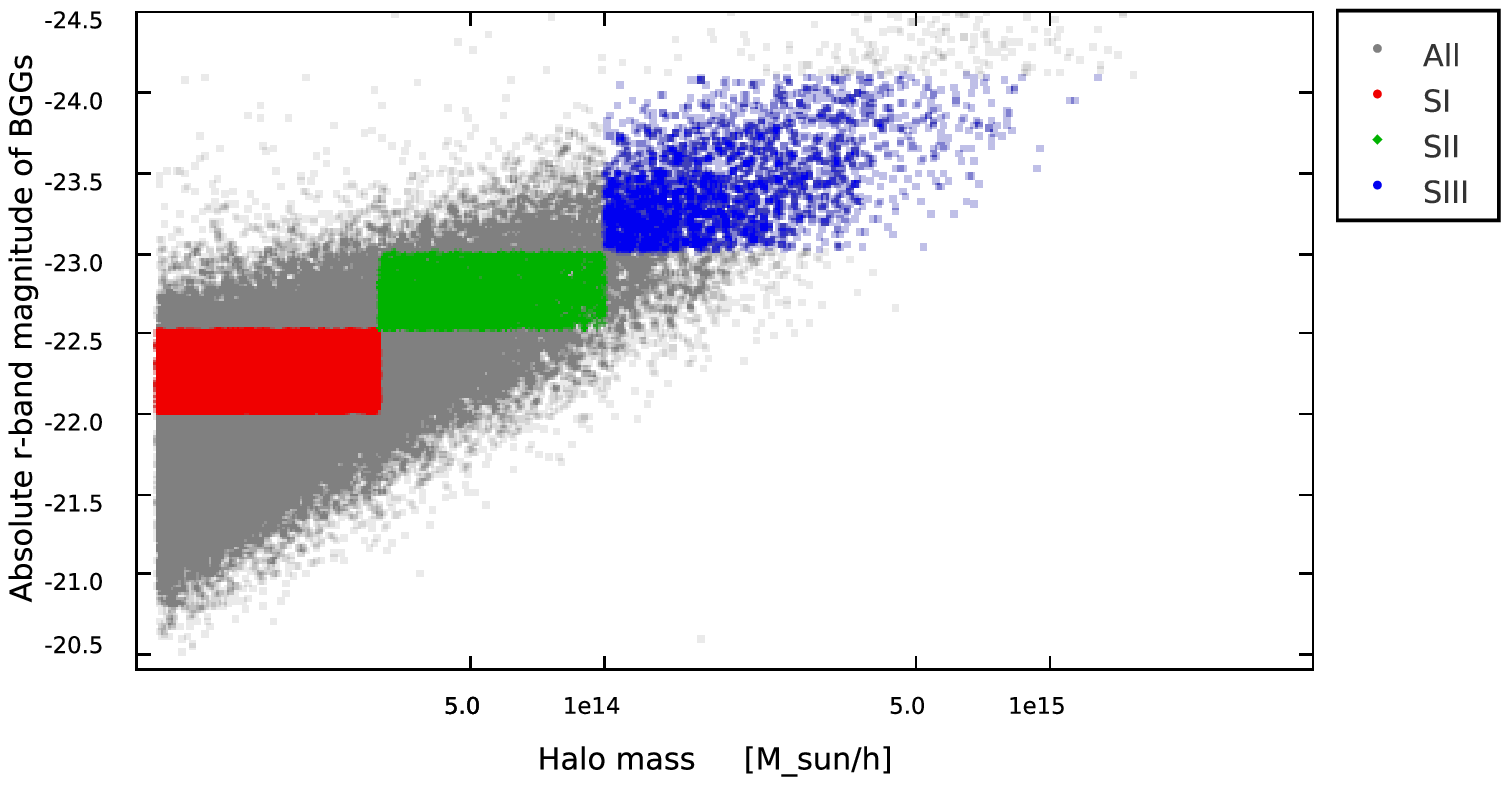}
      \end{center}
      \caption[halo_mass_bgg]{{\it Upper panel.} Magnitude gap
        distribution of all groups (filled gray histogram ) and groups
        selected according to the halo mass and BGG luminosity ranges of
        SI (solid red histogram), SII (dashed green histogram), and SIII
        (long-dashed blue histogram) at $ z=0 $ as described in \S 2.2. {\it
          Lower panel.} The BGG absolute r-band magnitude versus
        $M_{200}$ for all groups (gray points) at $ z=0 $. We highlight the sub-sample of SI, SII, and SIII with red, green, and blue points, respectively. The fossil, control, 
        and random groups are selected from these highlighted three sub-samples. }
      \label{mass_Lum}
      \end{figure*}
\cite{donghia05} found that fossils have assembled half of their
current dark matter mass at above z$\sim 1 $, having enough time to
merge L$^{*} $ galaxies and increase the magnitude gap between the
first and second brightest group galaxies. Using the Millennium
simulations, \cite{dariush07} revealed that fossils have accumulated a
larger fraction of their final halo mass at any redshift earlier than
non-fossil groups of similar X-ray luminosity, and concluded that an early
formation of fossil systems.  However, \cite{vonbendabeckmann08}
traced fossil groups since z$ \sim $1 and pointed out that the
majority of these systems fill their large magnitude gap due to the infall
of luminous galaxies, suggesting that the fossil phase is an
evolutionary stage of a galaxy group. Supporting this suggestion, we
show that some fossil groups in our XMM-CFHTLS galaxy groups catalogue
at $0.6<z<1.2$ include few galaxies outside the 0.5$R _{200}$, which are
brighter than the second brightest galaxy used for magnitude gap
calculation \citep{Gozaliasl14}.

\cite{Eigenthaler13} determined flat metallicity gradients for six
fossil galaxies, in contrast to the steep slope predicted by
a monolithic collapse, suggesting multiple mergers of galaxies for the
formation of fossil groups. Recently, following the fossil group
origins project, \citet{Aguerri11}, \citet{Zarattini14}, and
\citet{Girardi14} carried out a multi-wavelength study of a sample of
34 fossil group candidates, revealing that the brightest group galaxy (BGG) in a fossil group
forms from the merger/cannibalism of the $ L^{*} $ galaxies and the
large magnitude gap in fossil groups is the result of an evolutionary
effect and the extreme merger ratio of galaxies within these
systems. Moreover, \citet{Miraghaei14} have reported on the radio
properties of the brightest group galaxies in fossil groups in which they
discover that such galaxies are under-luminous in radio, compared to
non-fossil group galaxies of similar stellar mass.
  
We aim to use the semi-analytic model (SAM) of \citet[][hereafter G11]{Guo11}, to study
the composite LF of galaxies in fossil and non-fossil groups down
to an absolute magnitude of $ -15 $ mag. We use a well-defined, mass-selected sample of galaxy groups and clusters at redshift zero and
trace these halos and their properties up to $ z=1 $. In order to find any
effect of galaxy mergers on the composite LF of galaxies in fossil groups, we
examine evolution of the luminosity of the bright end and the faint
end slope as a function of redshift within a redshift range of 0$<z<$1. We compare our results with those of non-fossil groups with
similar halo mass and BGG luminosity. In this paper, we also investigate the
halo mass assembly history and the growth of the magnitude gap.

In \S 2, we describe data and define galaxy groups sample. Section 3 reports
the halo mass assembly history and evolution of the magnitude gap of
groups. Section 4 describes the galaxy LF and the redshift evolution of
Schechter function parameters. In \S 5, we discuss our results and present
conclusions.  Throughout this paper, we adopt a `concordance'
cosmological model, with $(\Omega_{\Lambda}, \Omega_{M}, w, h) =
(0.75, 0.25,-1, 0.73)$.

\section{Data and sample selections}

\subsection{Millennium simulation and semi-analytic galaxy catalogue}

The SAM of galaxy formation used in 
G11, is implemented on the stored output of a very large cosmological
N-body simulation, the Millennium simulation \citep{Springel05}.  The
Millennium simulation assumes a $\Lambda$CDM cosmology with
parameters, ($\Omega_{m}, \Omega_{b}, \Omega_{\Lambda}, n, \sigma_{8},
h) = (0.25, 0.045, 0.75, 1, 0.9, 0.73)$, based on a combined analysis
of the 2dFGRS \citep{Colless01} and the first-year WMAP data
\citep{Spergel03}.  Despite the fact that these values not the most recent
results \citep[e.g., WMAP7 and ][]{Planck13}, they correctly reproduce
the group abundance and clustering of dark halos at low-$ z $. The $
\sigma_{8} $ value is higher in the Millennium simulation than in the
WMAP7 results. This generally results in a discrepancy in the size and
the luminosity density of the structures \citep{Tavasoli13}. While
this might have an effect on the space density of fossil groups, the
present study focuses on the evolution of the composite luminosity
function based on SAMs, between $ z=1 $ and the present
day. \cite{Guo13} implement SAM on both Millennium and Millennium II
simulations in two cosmologies based on WMAP1 and WMAP7 to investigate
how the formation, evolution, and clustering of galaxies vary in
different cosmologies. The increased matter density
and decreased linear fluctuation amplitude $ \sigma_{8} $ in WMAP7
have compensating effects, so that the abundance and clustering of
dark halos are predicted to be very similar to those in WMAP1 for $z
\leqslant $ 3. Therefore, the main findings of our study should remain
intact.
 
The Millennium simulation traces $2160^{3}$ particles with typical
masses of $ 1.18 \times10^{9}$ M$_{\odot} $ from redshift 127 to the
present day. A friends-of-friends (FOF) algorithm is used to identify
groups by linking particles with separation less than $ \sim 0.2 $ the
mean inter-particle separation \citep{Davis85}. Identification of the
sub-halos in each FOF group is done using the SUBFIND algorithm
\citep{Springel01}.  The G11 is an updated and extended version of
earlier galaxy formation models based on the Millennium simulation
\citep{Springel05,Croton06,De Lucia07}. To estimate photometric
properties of galaxies, G11 uses a Chabrier IMF and stellar population
synthesis model from \cite{Bruzual03}.
        
Following \cite{white91}, G11 defines a cooling radius as the radius
within which the cooling time of the gas equals the current age of
the halo, and compares it with the virial radius ($ R_{vir} $).  If
the virial radius becomes smaller than the cooling radius, the gas is
accreted on the central galaxy on a free-fall timescale (rapid infall
regime). Conversely, if the virial radius becomes larger than the
cooling radius, cooling occurs in a statistic cooling flow.  G11
utilizes a new prescription for satellite galaxies: when a galaxy
passes outside $ R_{vir} $, it is no longer subject to the tidal and
ram-pressure forces. The gas stripping from satellites is gradual,
allowing the star formation to continue. Thus, in this model,
satellite galaxies can stay blue for longer periods. G11 also
includes starbursts triggered by a major merger associated with the
mass ratio of the merging progenitors in excess of 0.3. When a major
merger occurs, all of the stellar masses of the merging galaxies are
transformed to the bulge, while only a fraction of cold gas is turned
into the bulge stars, in accordance to the implementation of
\cite{Somerville01}. This model also includes the satellite-satellite
mergers and the disruption of satellite galaxies in presence of strong
tidal forces, which can reduce the number of satellite and dwarf galaxies
(which are generally overpredicted by the most SAMs) in massive
halos. G11 uses a new disc model that separately allows the gas
and stellar discs to grow continuously in mass and angular momentum.
A velocity dependent model of supernovae feedback model is adopted in
G11. G11 model also implements both radio and quasar modes of the
active galactic nuclei (AGN) feedback following \cite{Croton06}. It
also allowed the radio mode of AGN feedback to occur in satellite
galaxies.

\begin{figure}
  \begin{center}  
    \leavevmode
  \includegraphics[width=8.5cm]{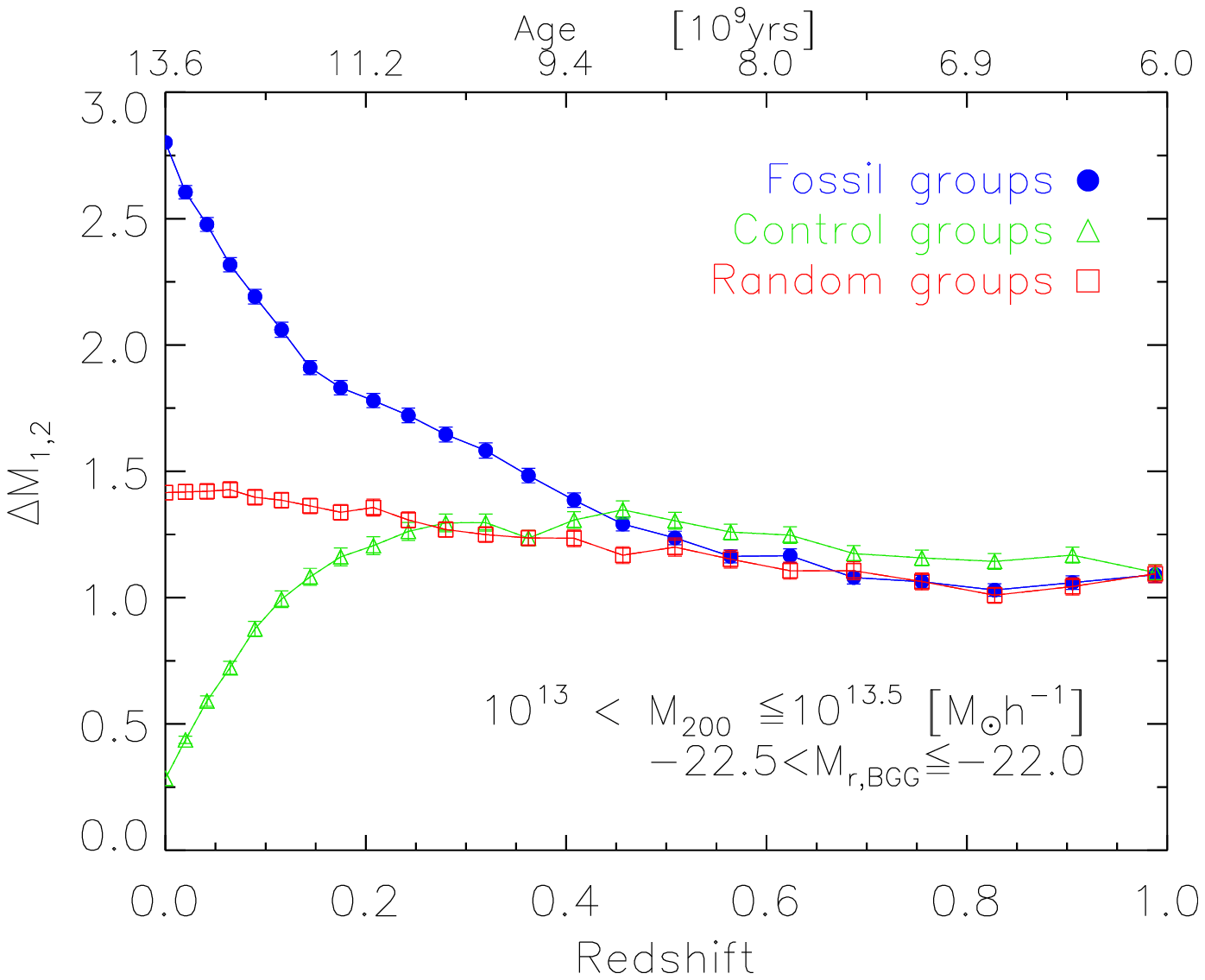}
  \includegraphics[width=8.5cm]{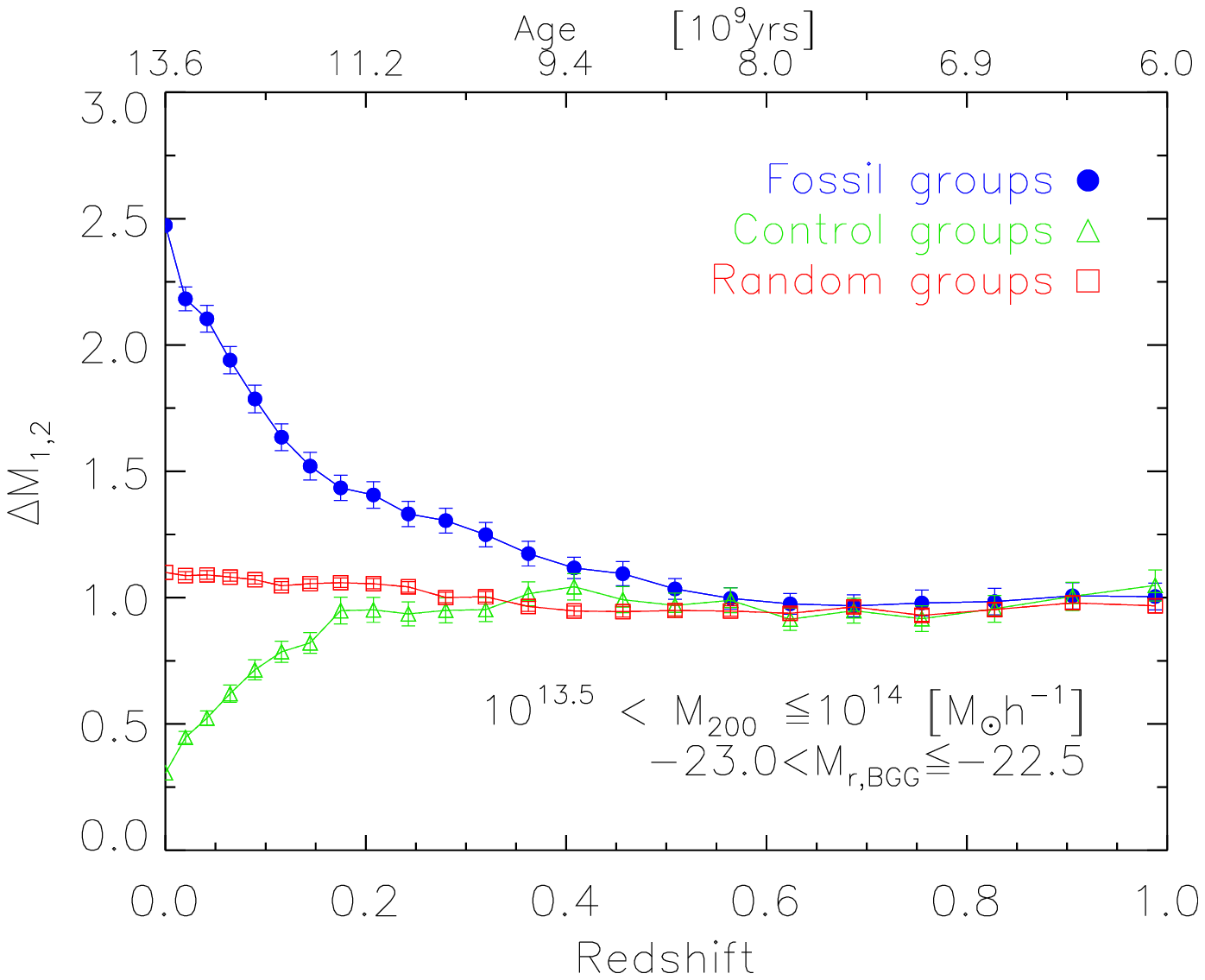}
  \includegraphics[width=8.5cm]{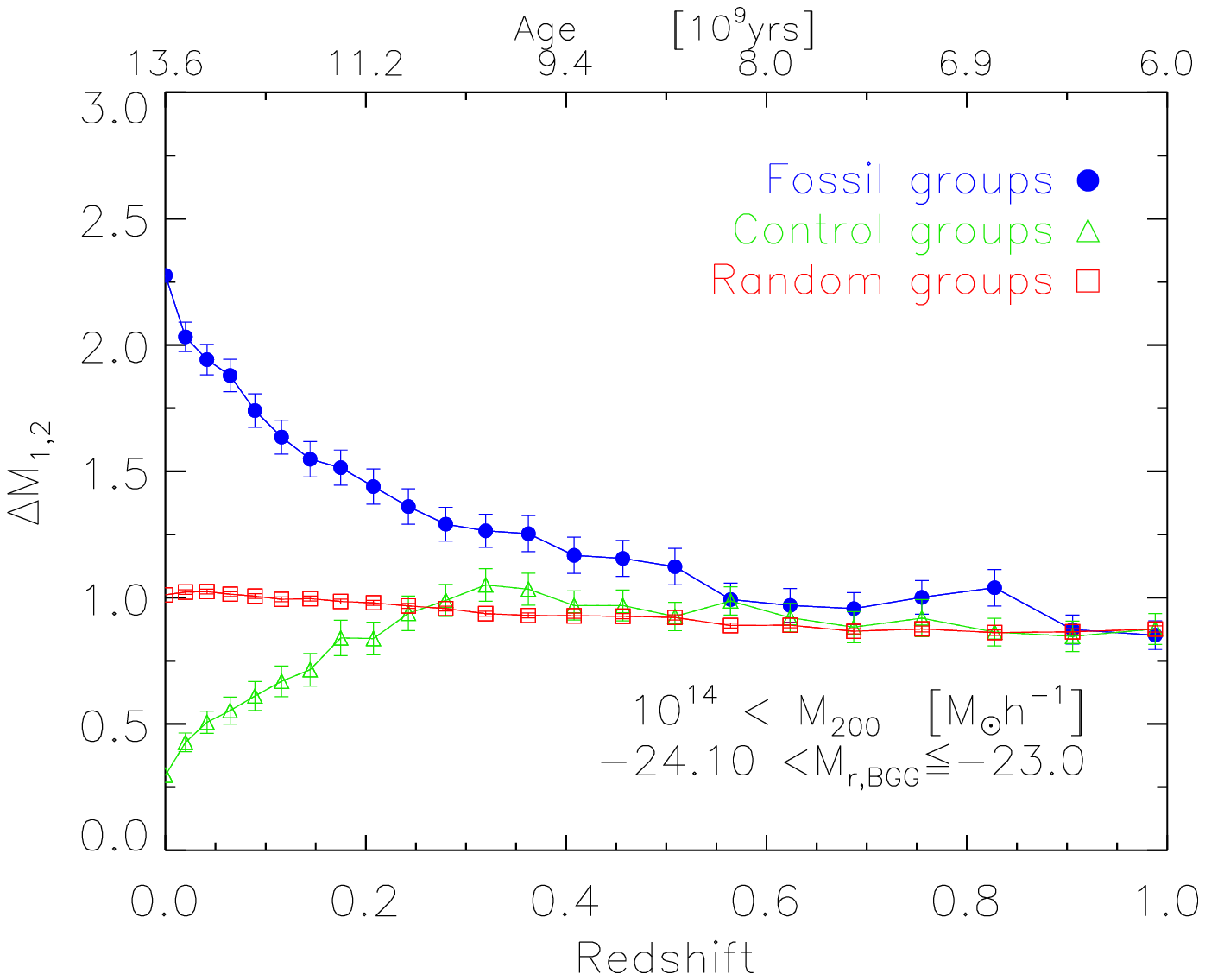}
 
       \end{center}
       \caption[flag]{Evolution of the luminosity gap from $z=0$ to $z=1$
         for fossil, control, and random groups. The filled blue
         circles, open green triangles, and open red squares show the
         mean $\Delta M_{1,2}$ as a function of redshift for fossil,
         control, and random groups, respectively, within SI (top
         panel), SII (middle panel), and SIII (bottom panel). The
         significant formation of magnitude gap occurs at redshifts
         $0<z<0.5$.  }
     \label{trace_delta}

\end{figure}

 \begin{figure}
   \begin{center}  
     \leavevmode
   \includegraphics[width=8.5cm]{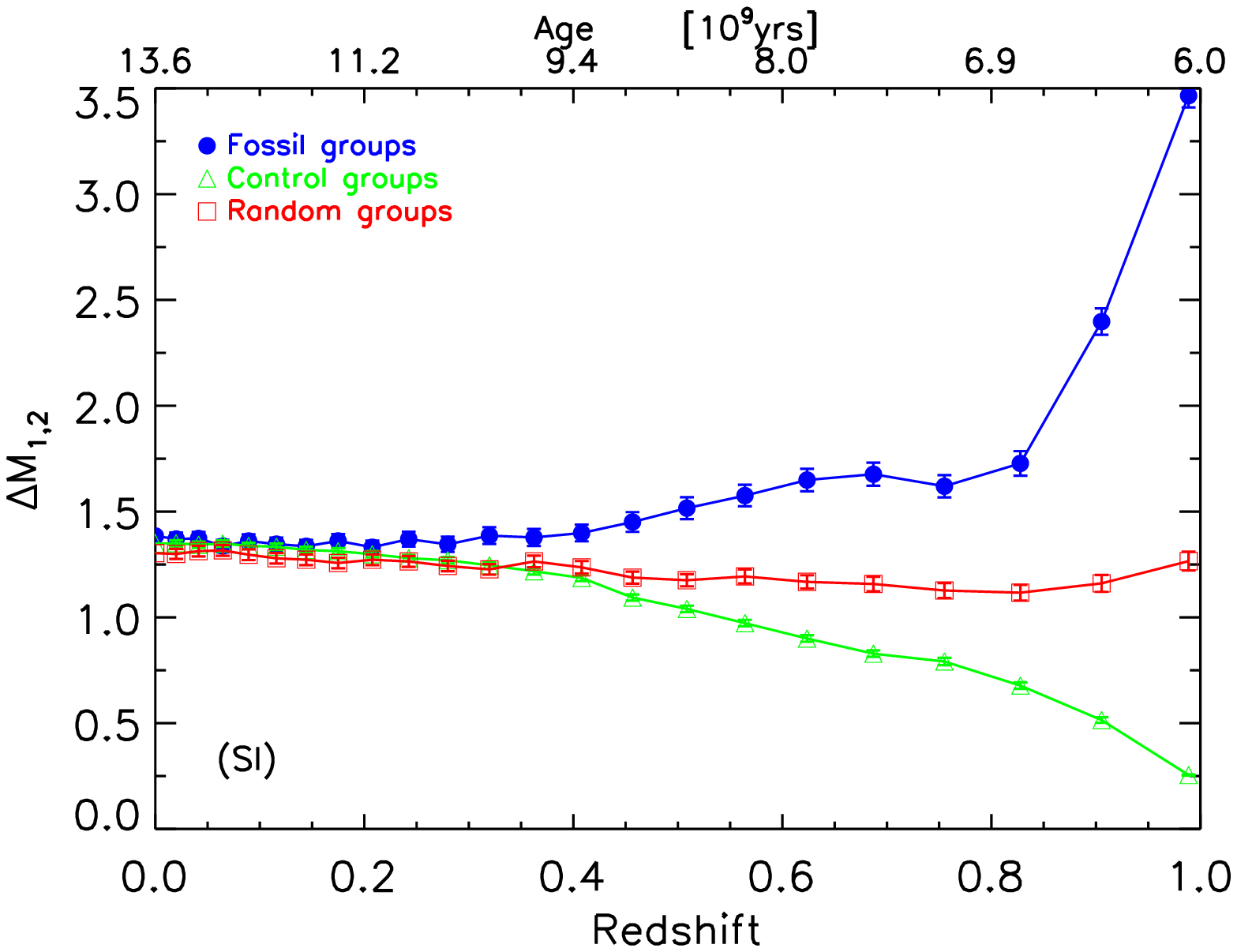}
   \includegraphics[width=8.5cm]{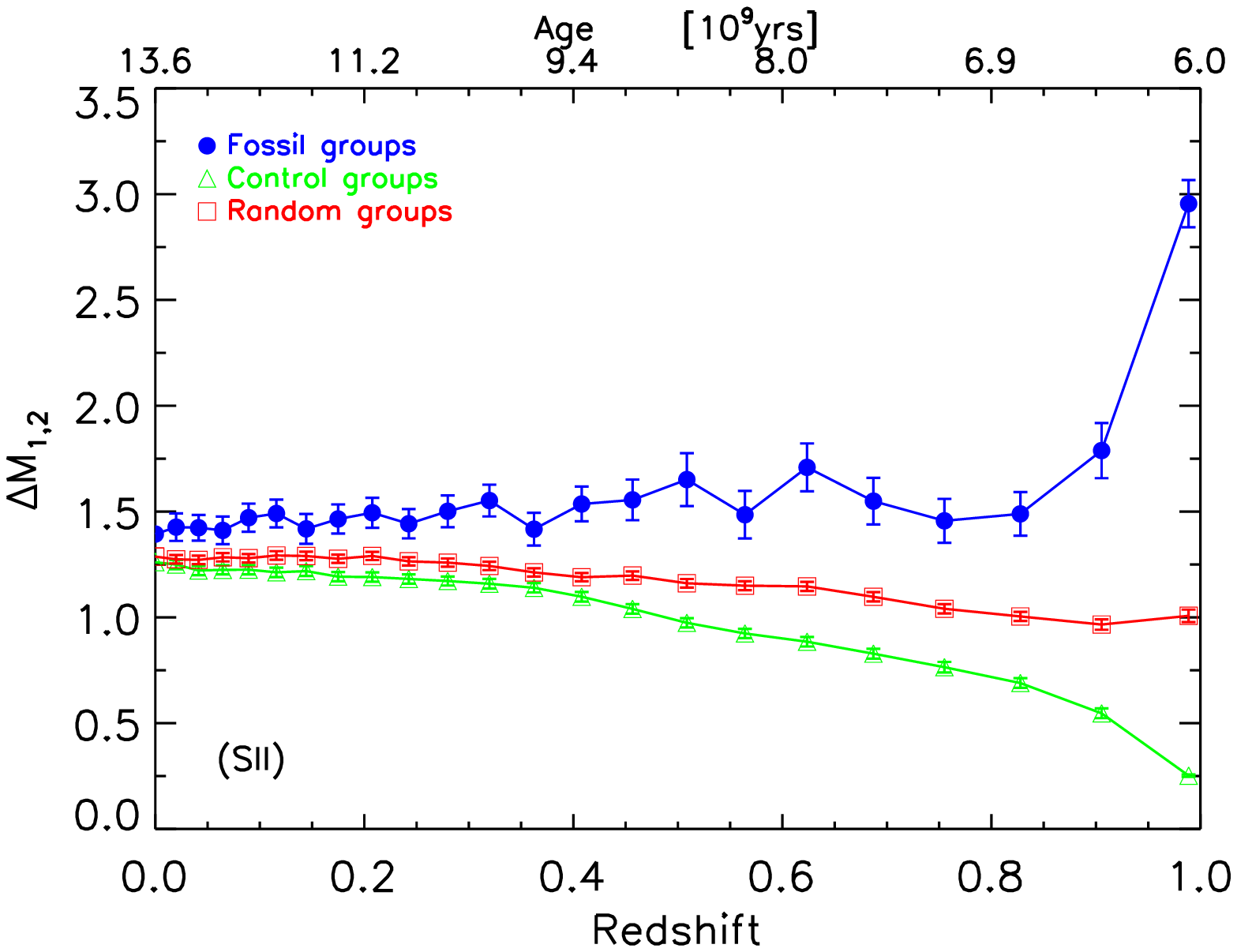}
   \includegraphics[width=8.5cm]{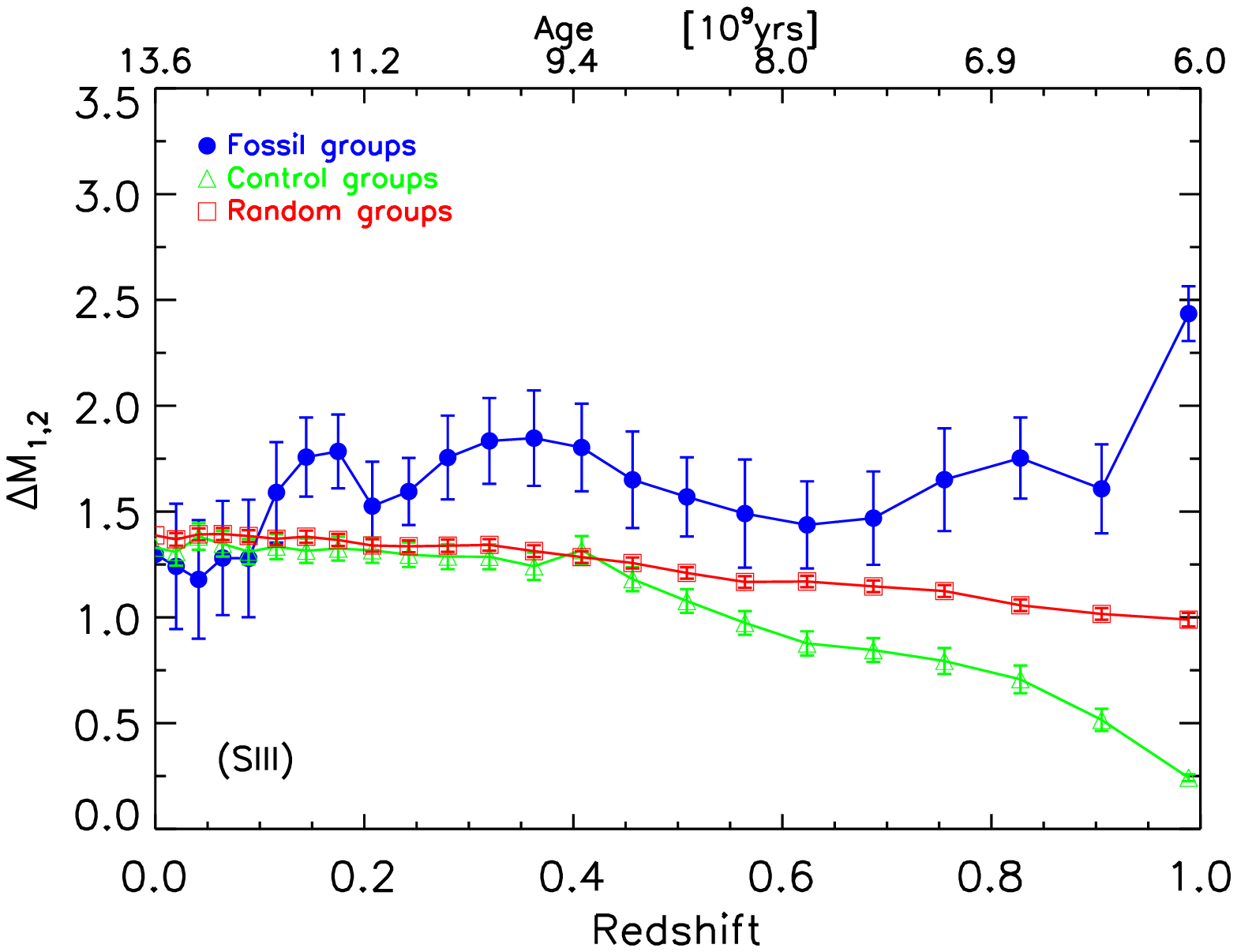}
  
        \end{center}
        \caption[flag]{As in Fig. \ref{trace_delta}, except for the redshift
          evolution of the mean magnitude gap for fossil, control, and
          random groups. The groups are traced forwards in time, from
          $ z=1 $ to $ z=0 $.  }
      \label{gap_inv}
 
 \end{figure} 
\subsection{Definition of fossil groups and sample selection}

In order to define our sample of galaxy groups (e.g. fossil groups) we
assume the coordinates of the BGGs as group centres and then measure
the magnitude gap, $\Delta M_{1,2}$, between the first and second
brightest galaxies within a half of the projected radius enclosing 200
times the mean density, $R_{200}$, of each galaxy group, similar to
the convention used in studies of fossil groups \citep{jones03}.  We define
three samples of galaxy groups based on the magnitude gap
measurements:

(i) groups  with $\Delta M_{1,2}\geq 2$ mag  as fossil groups,

(ii) groups with $\Delta M_{1,2}\leq 0.5$ mag as non-fossil or control groups, and

(iii) to understand whether $\Delta M_{1,2}$ distinguishes between
galaxy groups, we also select a sample of groups with any values of
$\Delta M_{1,2}$. This helps us find the robustness of $\Delta
M_{1,2}$ as an optical criterion in identifying fossil groups.
 
The constraint on the halo mass results in an initial sample of 51366
galaxy groups at z$ = 0$, with 13 per cent fossil groups and
25 per cent $\Delta M_{1,2}\leq 0.5$ mag  control groups.

For each sample of galaxy groups, we define three sub-samples
following the halo mass and the BGG luminosity ranges:
 
(SI)  $13.0<log(M_{200}/h^{-1}M_{\odot})\leq13.5$  and $ -22.5<M_{r,BGG}\leq -22$, 
 
 (SII) $ 13.5<log(M_{200}/h^{-1}M_{\odot})\leq14.0$ and $  -23 <M_{r,BGG}\leq -22.5 $, and  
   
  (SIII) $ 14.0<log(M_{200}/h^{-1}M_{\odot})$ and  $ -23.0 <M_{r,BGG}\leq -24.10 $.
  
  We select 1481, 307, and 89 galaxy groups at $z=0$ of each group type
  obeying the SI, SII, and SIII criteria, respectively. The constraint
  on the luminosity of BGG enables us to select relatively homogeneous
  sub-samples and also to detect a presence of a large
  magnitude gap in the shape of composite LF of fossil
  groups. Otherwise, the large gap in the bright end of the composite
  LF is filled by combining groups with different BGGs. In
  Fig. \ref{mass_Lum} (lower panel), we plot the absolute r-band
  magnitude of BGGs ($M_{r,BGG}$) versus halo mass for all groups
  (gray points) and for all group sub-samples at $z=0$. We select our sample of the fossil, control, and random groups from the  highlighted sub-samples.
\begin{figure}
  \begin{center}  
    \leavevmode
  \includegraphics[width=8.5cm]{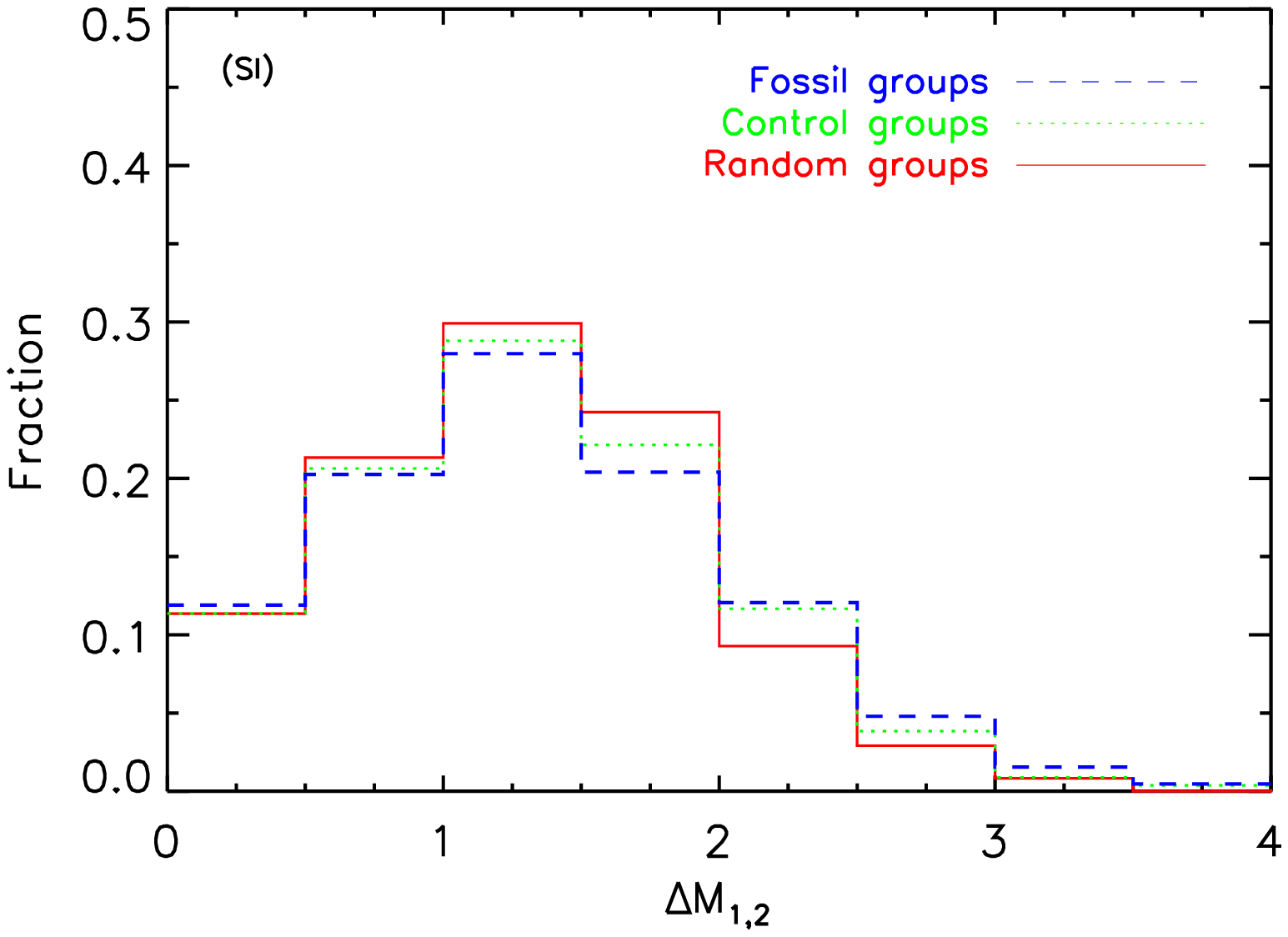}
  \includegraphics[width=8.5cm]{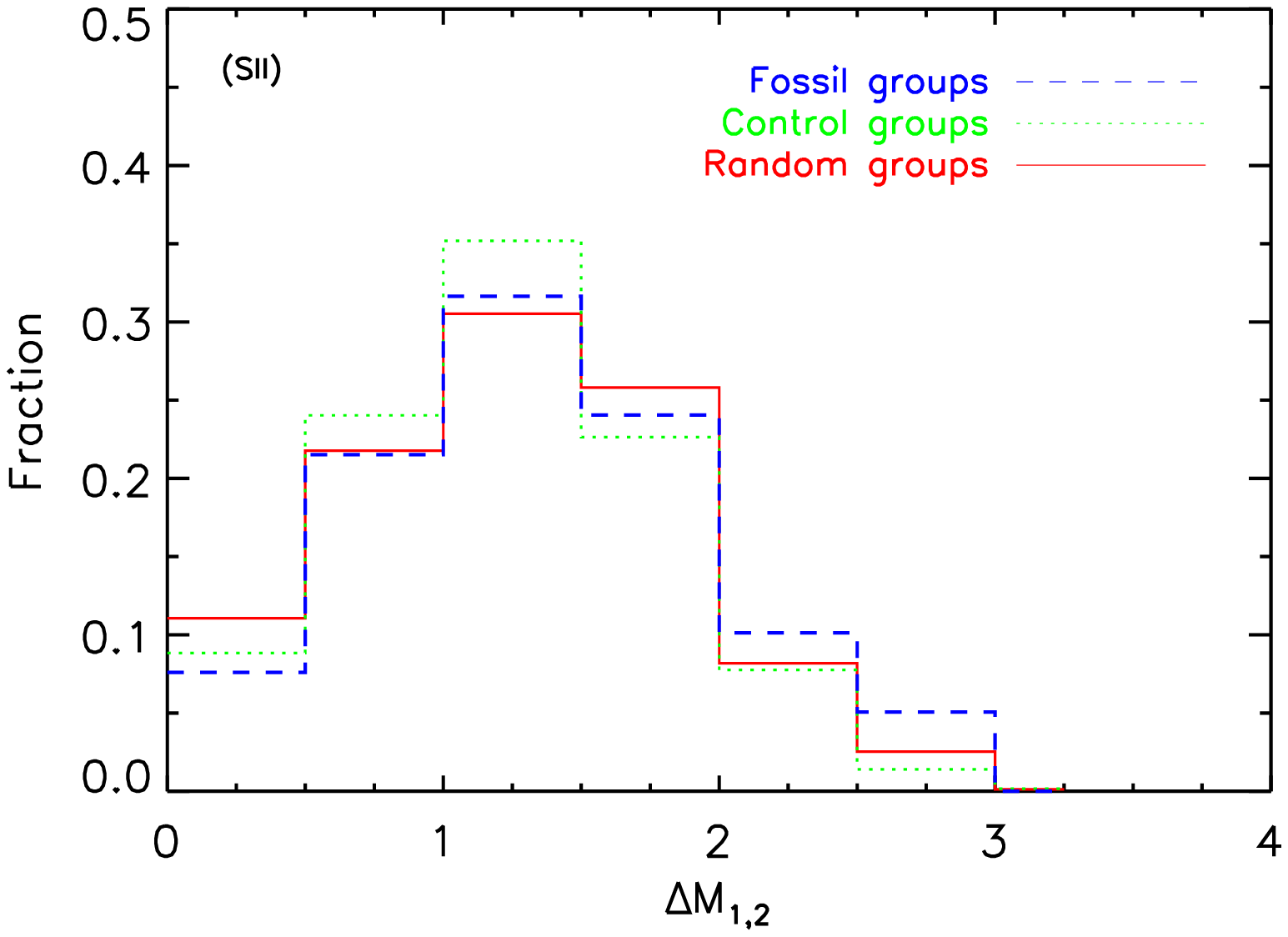}
  \includegraphics[width=8.5cm]{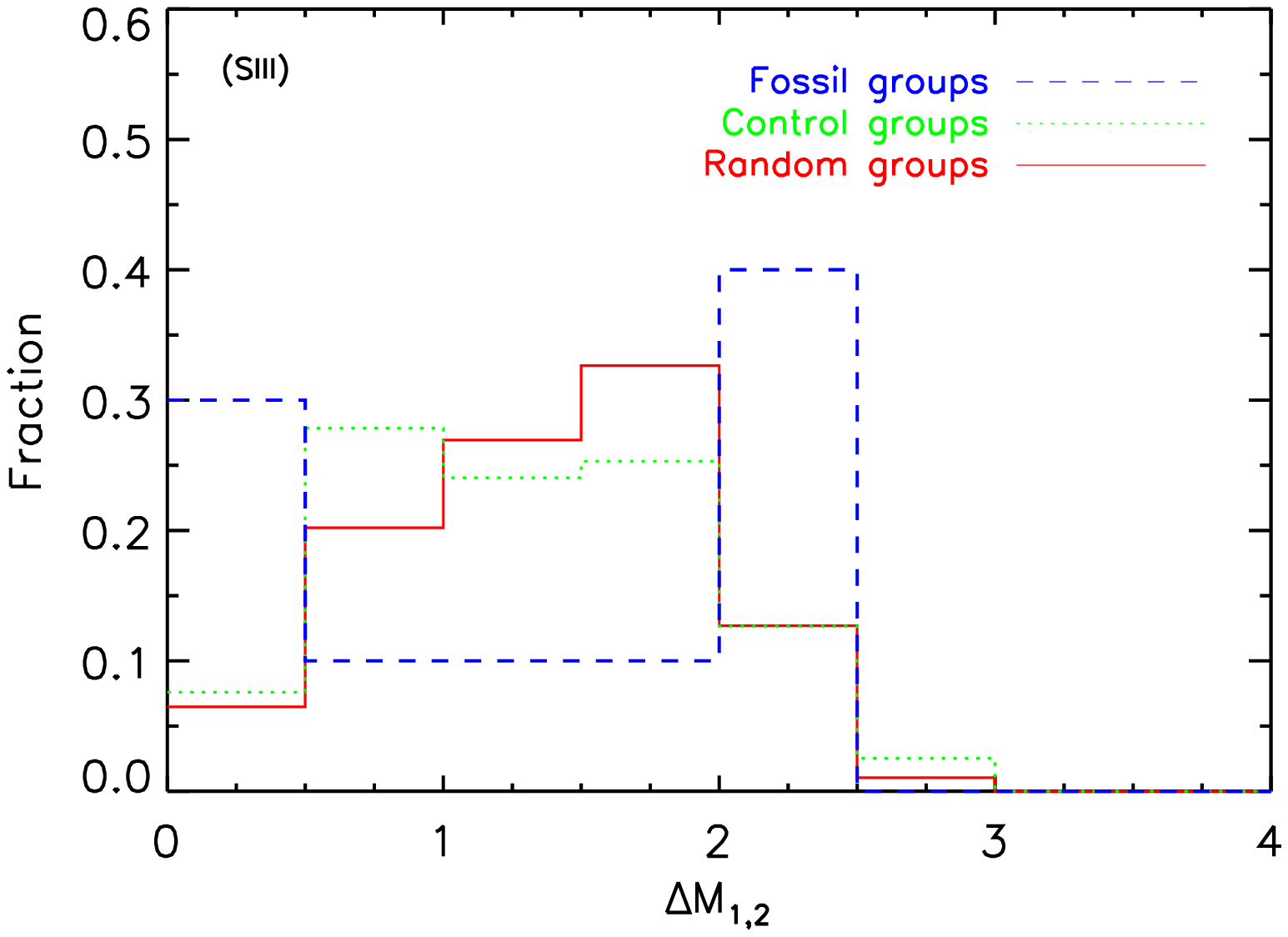}
        \end{center}
        \caption[flag]{Magnitude gap distribution of the remnant
          groups at $ z=0 $ for the halos that are selected as fossil, control, 
          and random groups at $ z=1 $ and traced to $ z=0 $. The blue dashed,
          green dotted and red solid histograms show a gap distribution
          for traced fossil, control, and random groups within SI
          (top panel), SII (middle panel), and SIII (bottom panel). It
          appears that $ \approx40\% $ of fossils with $ M_{200}>
          10^{14} M_{\odot}h^{-1} $ survive their gap after evolving
          between $ z=1 $ and $ z=0 $. }
     \label{gapdis_inv}

\end{figure}
\section{ The $\Delta M_{1,2}$ and the halo mass evolution of galaxy groups}
\subsubsection{ Tracing the $\Delta M_{1,2}$ backwards }

Magnitude gap has been often used as a tool to understand the
dynamical age and formation history of galaxy groups
\citep{Milosavljevic06,dariush07,dariush10,vandenBosch07}. Based on
simulations, \cite{donghia05} showed that  early formed galaxy
groups, such as fossil groups, have a large magnitude gap. In Fig.
\ref{mass_Lum} (upper panel), we show the magnitude gap distribution
for all groups at z=0 and for SI (solid red histogram), SII (dashed
green histogram), and SIII (long-dashed blue histogram). It appears
that the fraction of fossil groups decreases with increasing halo
masses.

In order to understand the time evolution of $\Delta M_{1,2}$
parameter, we trace backwards all halos in our sub-samples from z$=0 $
to z$=1$ and measure their mean magnitude gaps at each redshift
slice. In Fig.\ref{trace_delta}, we illustrate the mean magnitude gap
as a function of redshift for SI (top panel), SII (middle panel), and
SIII (bottom panel) sub-samples. We find that the mean magnitude gap
of fossils decreases to a lower value with increasing redshift, so that
at z$ \sim $0.5 we detect no differences between all three samples of
groups. In contrast to fossils, the mean magnitude gap for the control
sample grows with increasing redshift between $ z=0 $ to $ z=0.5 $. We detect
no redshift evolution for the mean magnitude gap of random
groups. We detect no difference for the magnitude gap evolution between different subsamples shown in   Fig. \ref {trace_delta}.

\begin{figure}
  \begin{center}  
    \leavevmode
  \includegraphics[width=8.5cm]{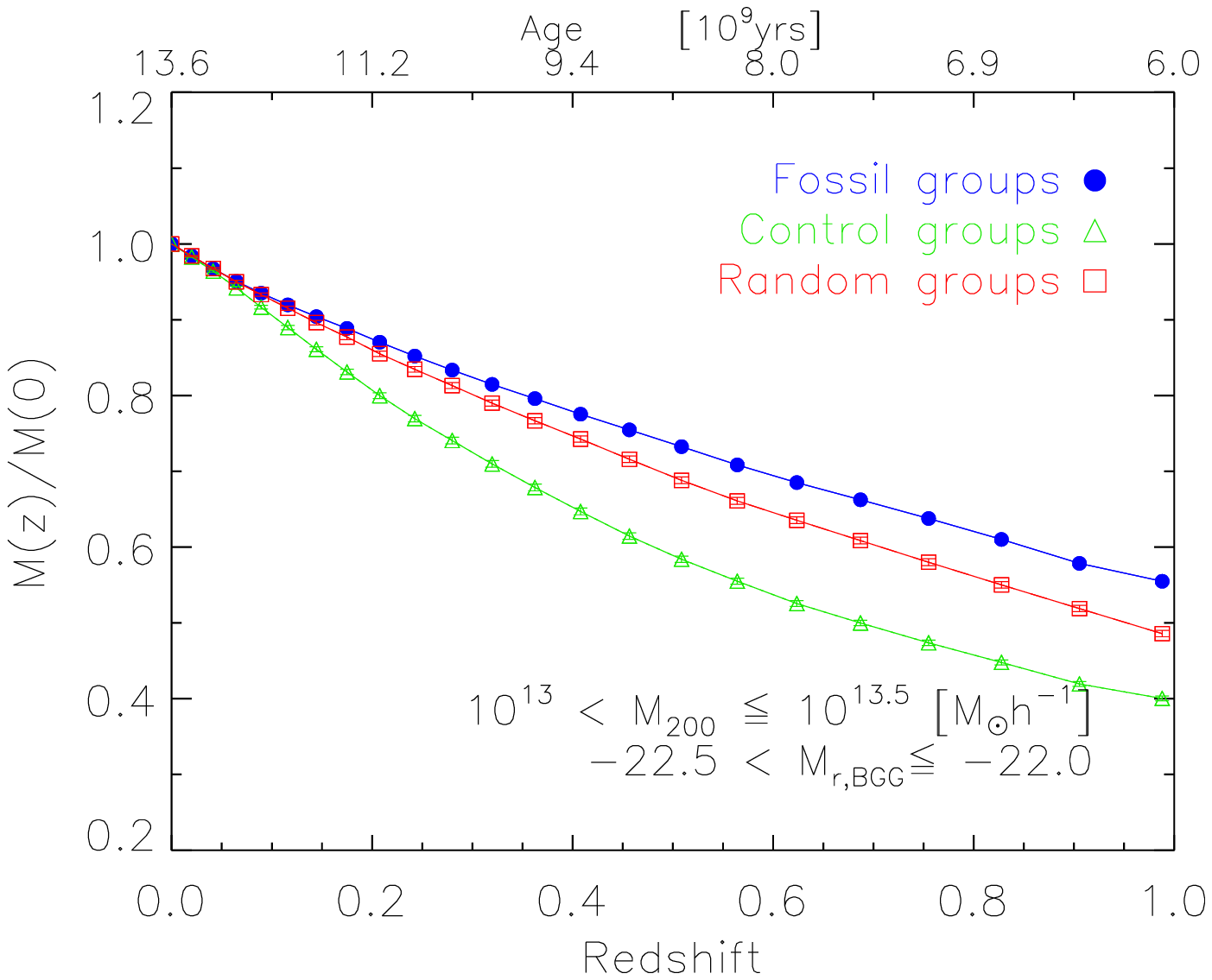}
  \includegraphics[width=8.5cm]{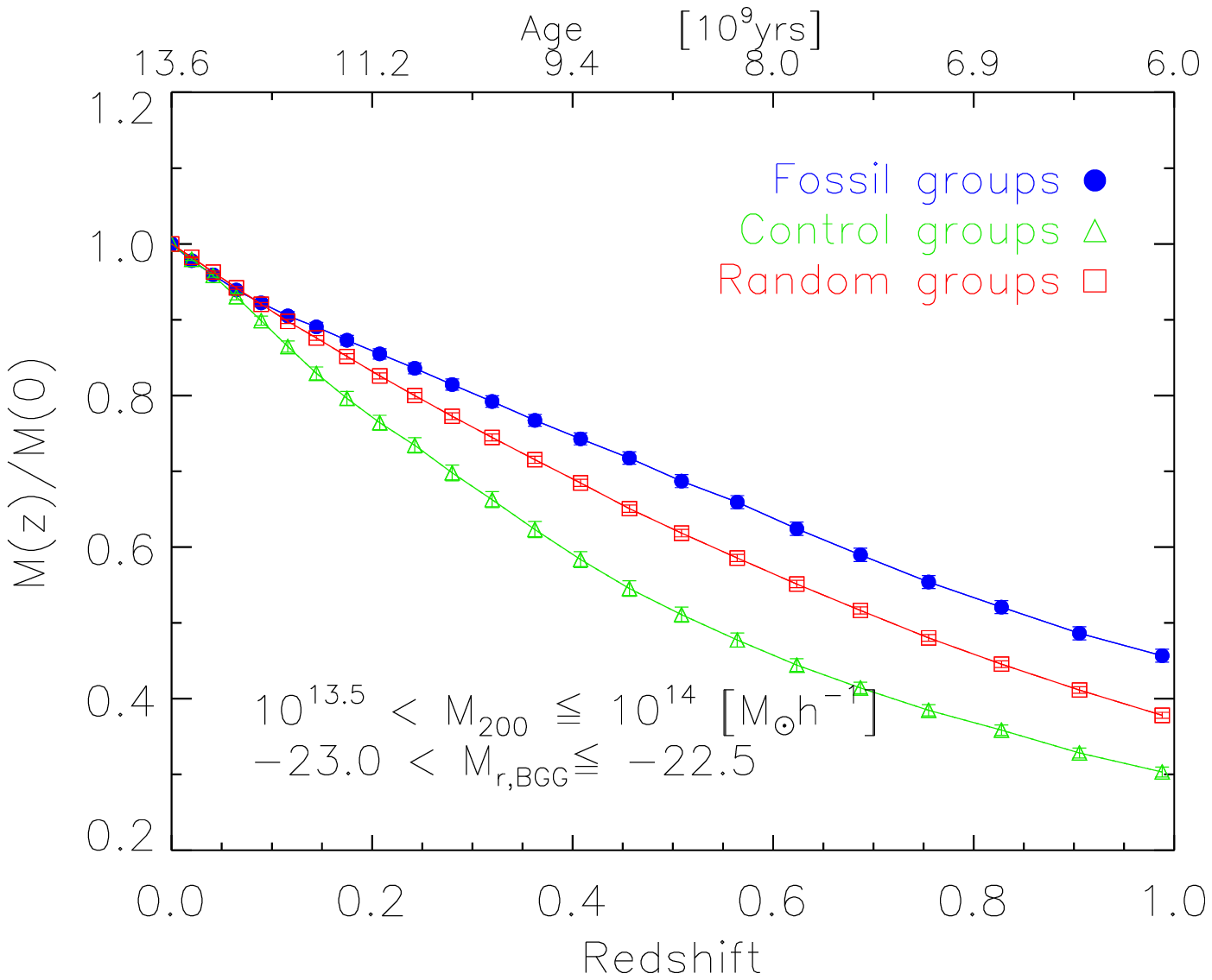}
  \includegraphics[width=8.5cm]{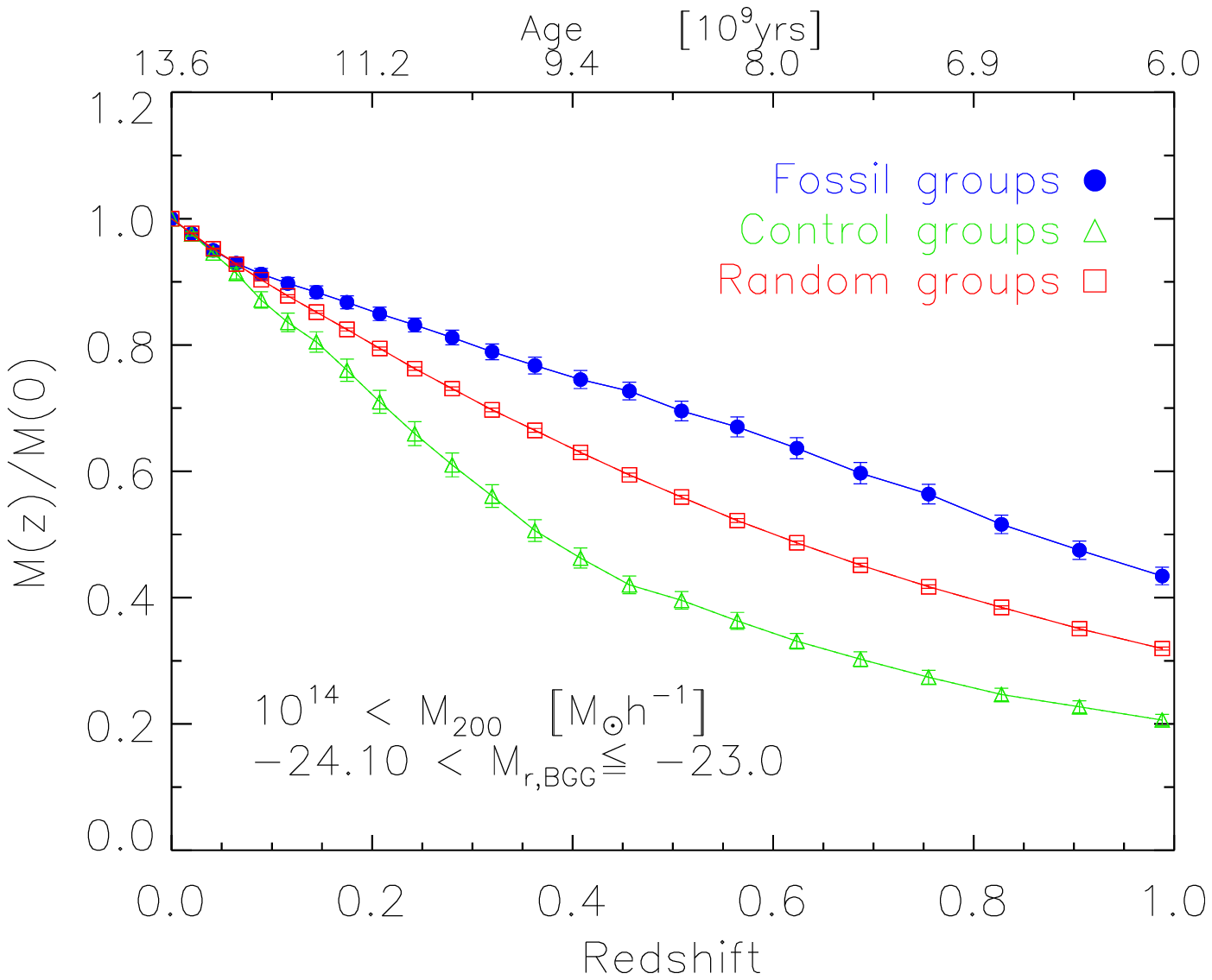}
 
       \end{center}
       \caption[flag]{Tracing back the mass assembly of the dark
         matter halos as a function of Hubble time and redshift for
         fossil, control, and random groups for SI (top panel), SII
         (middle panel), and SIII (bottom panel) samples. }
     \label{t_mass}

\end{figure}

\subsubsection{Tracing  the $\Delta M_{1,2}$ forwards}
 
Similar to \cite{vonbendabeckmann08}, we examine whether the magnitude
gap of groups evolves forwards in time and select three different
sub-samples of fossil, control, and random groups at $ z=1 $, according to
the mass ranges defined in \S 2.2. We use a similar cut for the
absolute magnitude of BGGs ($ M_{r}\geqslant-22 $ mag) for all
halos. We track these groups forwards in time from z$=1 $ to z$=0 $
(Fig. \ref{gap_inv}). In agreement with \cite{vonbendabeckmann08}, we
find that the majority ($\sim$ 80 per cent) of fossil groups with a halo
mass less than $ M_{200}=10^{14} $ $ M_{\sun} $h$ ^{-1} $ fill their
large magnitude gaps (e.g., due to the infall of luminous galaxies
within $ 0.5R_{200} $, where we calculate the magnitude gap). However,
about 40 per cent of fossil clusters with $ M_{200}>10^{14} M_{\sun}
$h$ ^{-1} $ retain their large magnitude gap. In addition, the mean
gap value for the control groups increases with decreasing
redshift. For random groups, the mean gap value remains flat with
redshift. The mean gap value of all group types in SI becomes
similar below $z\sim $0.3, but for SII and SIII the mean gap values
become similar more recently, $z\sim $0. 

In Fig. \ref{gapdis_inv}, we show the magnitude gap distributions at
$ z=0 $ for the traced groups since $ z=1 $. We also find that $ \approx$ 15
per cent of control groups, which tracing from $ z=1 $, become fossil groups at $ z=0 $.

\section{Evolution of the halo mass }

Previous studies argued that fossil groups are early formed,
relatively old, and isolated systems \citep{dariush07,donghia05}.
Similar to other studies e.g. \cite{dariush07,dariush10}, we
investigate the halo formation history and mass assembly of galaxy
groups in our sub-samples of galaxy groups using the G11 model. In
Fig. \ref{t_mass}, we trace backwards the mass build-up of dark
matter halos as a function of redshift for fossil, control, and random
groups from  $ z=0 $ to $ z=1 $.

The vertical axis in Fig. \ref{t_mass}, represents a ratio of the mean
mass of selected halos at each redshift to their final mass at
z=0. The error bars are the standard error on the mean.  A similar
evolution is detected for the halo mass build-up of fossil groups
within SI, SII, and SIII. Fossils have assembled $\approx$50 per cent
of their current mean halo masses at z$=1 $, while controls and random
groups accumulate $ \approx$30 per cent and $\approx$40 per cent,
respectively. In agreement with early studies we find that fossils are
early formed groups compared to control groups with similar halo
masses and BGG luminosities. We also find that low-mass groups have
assembled larger fraction of their masses at $ z=1 $ compared to the high-mass groups. For example, fossils within SI, SII, and SIII are
accumulated $ \approx$57, 46, and 41 per cent of their final halo mass
at $ z=1 $, respectively. Fig. \ref{t_mass} also shows that the mean
halo mass of selected fossil, control, and random groups evolve with
different trends at $ 0.2 < z< 1 $. Similar trends are observed if
halos are traced forwards from $ z=1 $ to $ z=0 $.
 
 \begin{figure}
  \begin{center}  
    \leavevmode
      \includegraphics[width=8cm]{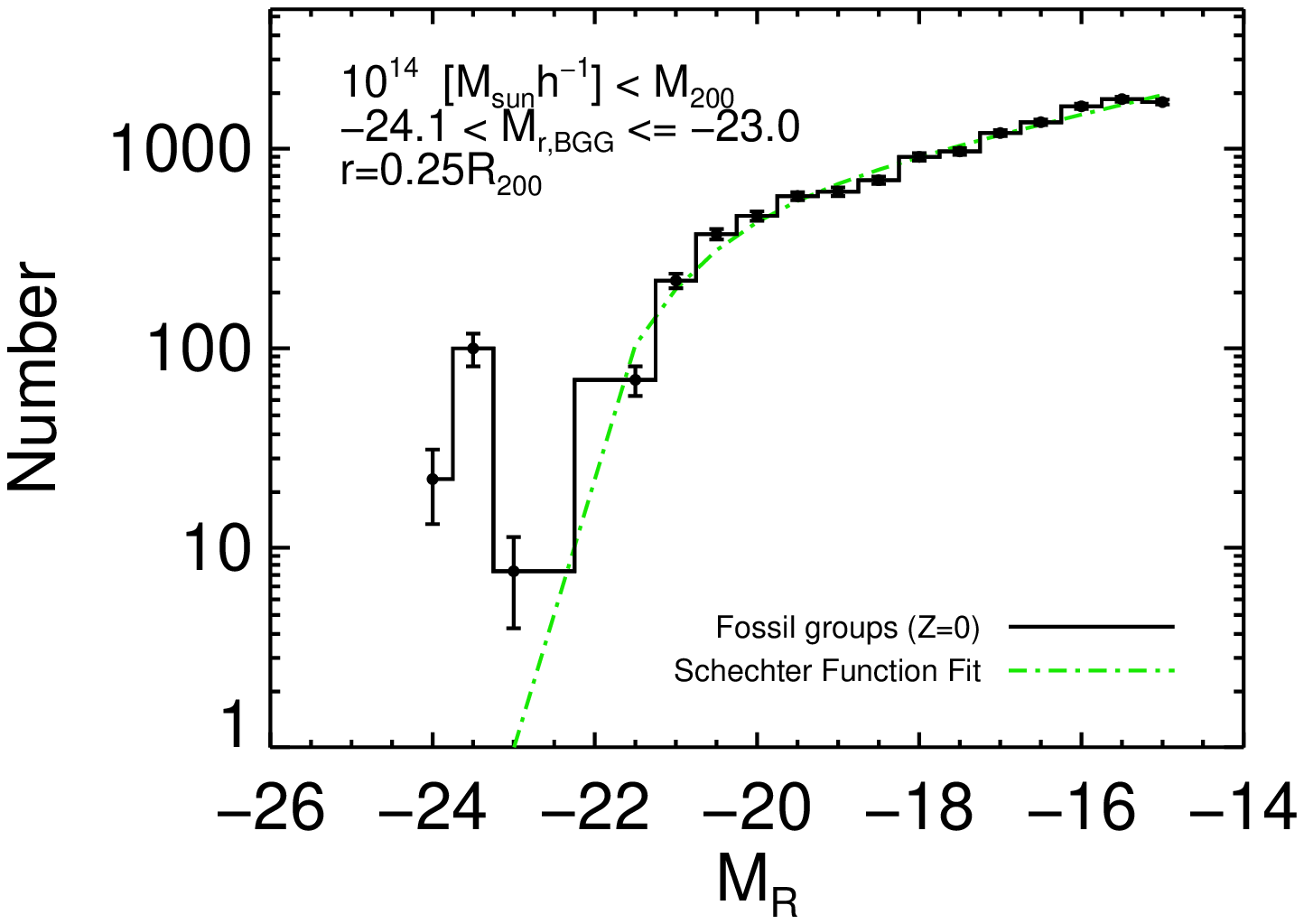}
  \includegraphics[width=8cm]{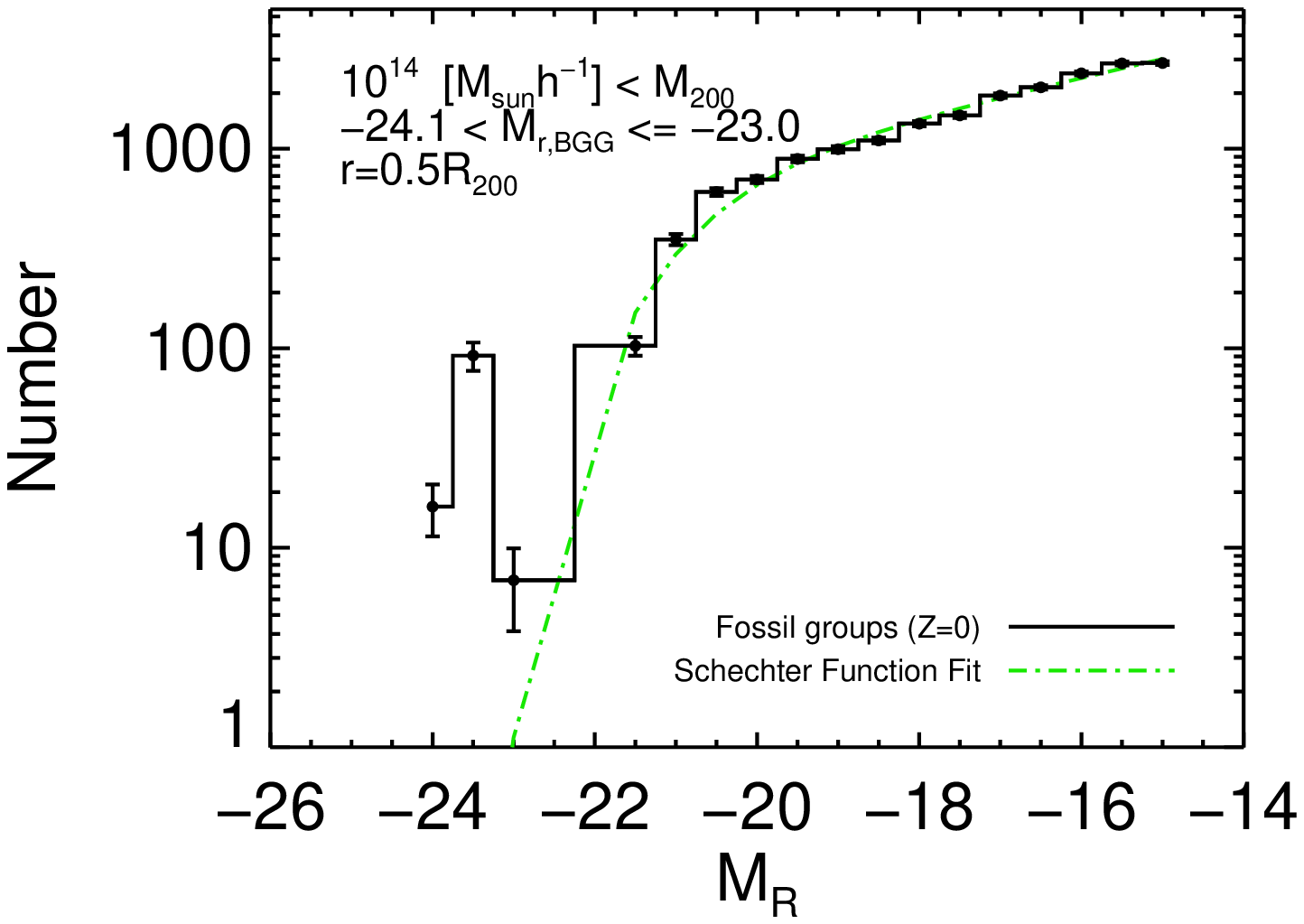}
    \includegraphics[width=8cm]{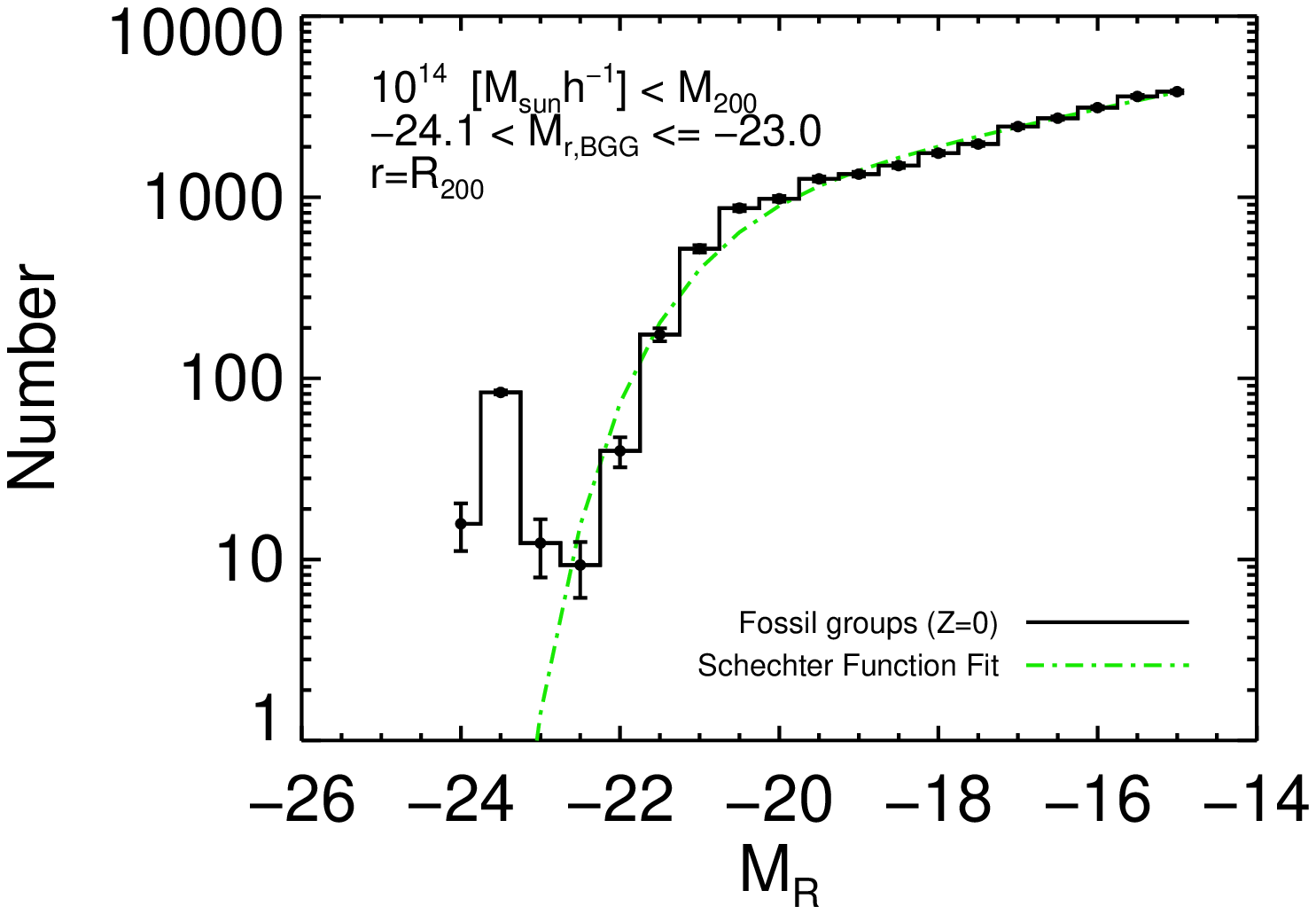}

       \end{center}
       \caption[flag]{The composite LFs of galaxies in fossil groups
         and the single Schechter function fit to the data (green
         curve) within the $ \frac{1}{4} R_{200} $ (top-left panel), $
         \frac{1}{2} R_{200} $ (top-right panel), and $ R_{200} $
         (bottom-left panel) for SIII.  }
     \label{LFs}
\end{figure}

\section{Luminosity Function}
\subsection{Composite luminosity function}
 
Luminosity function (LF) of galaxies is used as a key function to advance our understanding of galaxy properties in different
environments
\citep{Bahcall79,Binggeli88,Blanton01,Yang03,Benson03,Lin96,Milosavljevic06,Tinker09,vandenBosch07,Popesso05}.
As such, the observed LFs are regarded as fundamental observational
quantities and they must be reproduced by a successful theoretical
model of galaxy formation and evolution. A number of studies have
detected substantial differences between the LF of galaxies in
the field, group, and cluster environments that can arise from differing of the galaxy formation,  because of the differences in the galaxy
environments, or that can be related to effects on the LF by
dynamical physical processes (i.e. mergers, effects of tidal and
ram-pressure forces) that occur during or after the collapse of galaxy
groups/clusters \citep[e.g.,][]{Miles06}. Several earlier studies
attempted to show similarities between LFs in different environments,
suggesting the LF as universal function
\citep[e.g.,][]{Olmer74,Gaidos97,Colles89,Propris03}. In contrast, a
number of studies disagree with the universal shape of LF
\citep[e.g.,][]{Godwin77,Piranomonte01,Hansen05,Giodini12}.
 \begin{figure*}
  \begin{center}  
    \leavevmode
 
  \includegraphics[width=15cm]{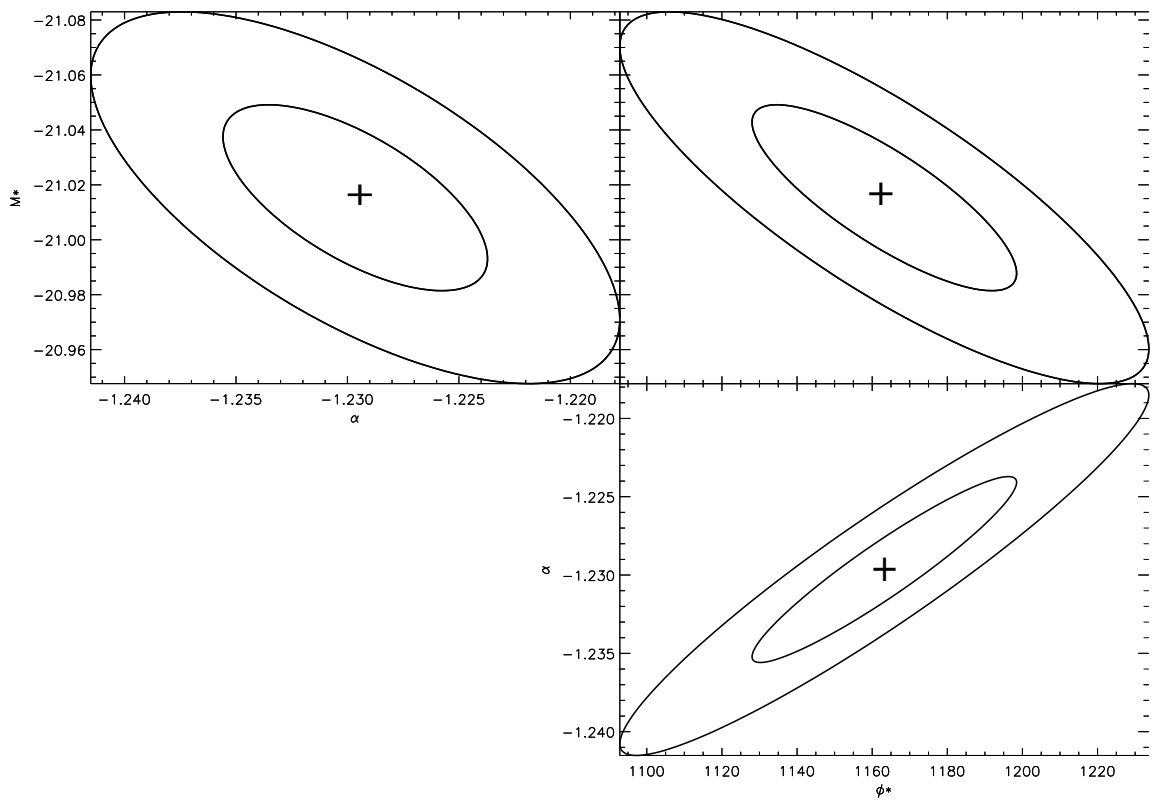}

       \end{center}
       \caption[flag]{ The 1-$ \sigma $ and $ 2-\sigma $ error
         contours for the fit on the composite LF of galaxies in
         fossil groups at $ z=0 $ for SIII. }
     \label{sigma}
\end{figure*} 
The luminosity evolution of the bright end and slope of the faint end
of the galaxy LF in the field and clusters has been well understood
\citep{Bowler14,Lilly95,Norberg02,Willmer06,Ellis96,Alshino10}. However,
because of a lack of data for fossil groups, the nature of such
evolution in fossils has remained an open question.

Using numerical simulation of the merging of a compact group,
\cite{barnes89}, concluded that an elliptical galaxy can be formed by
merging galaxies in a few tenths of a Hubble time. Later,
\cite{ponman93} and \cite{ponman94} studied HCG 62 and RX $
J1340.6+1018 $, finding that compact groups of galaxies form as a
result of the orbital decay of galaxies into an extended dark matter
halo. Finally, massive galaxies in these systems merge with the central group galaxy as a result of dynamical friction, forming a system known as a fossil group. Such a fossil group includes a giant elliptical galaxy with an extended hot gas halo that displays a large difference in magnitude relative to the second brightest satellite galaxy  within $ 0.5R_{200} $.

 \begin{figure}
   \begin{center}  
     \leavevmode
   \includegraphics[width=8.5cm]{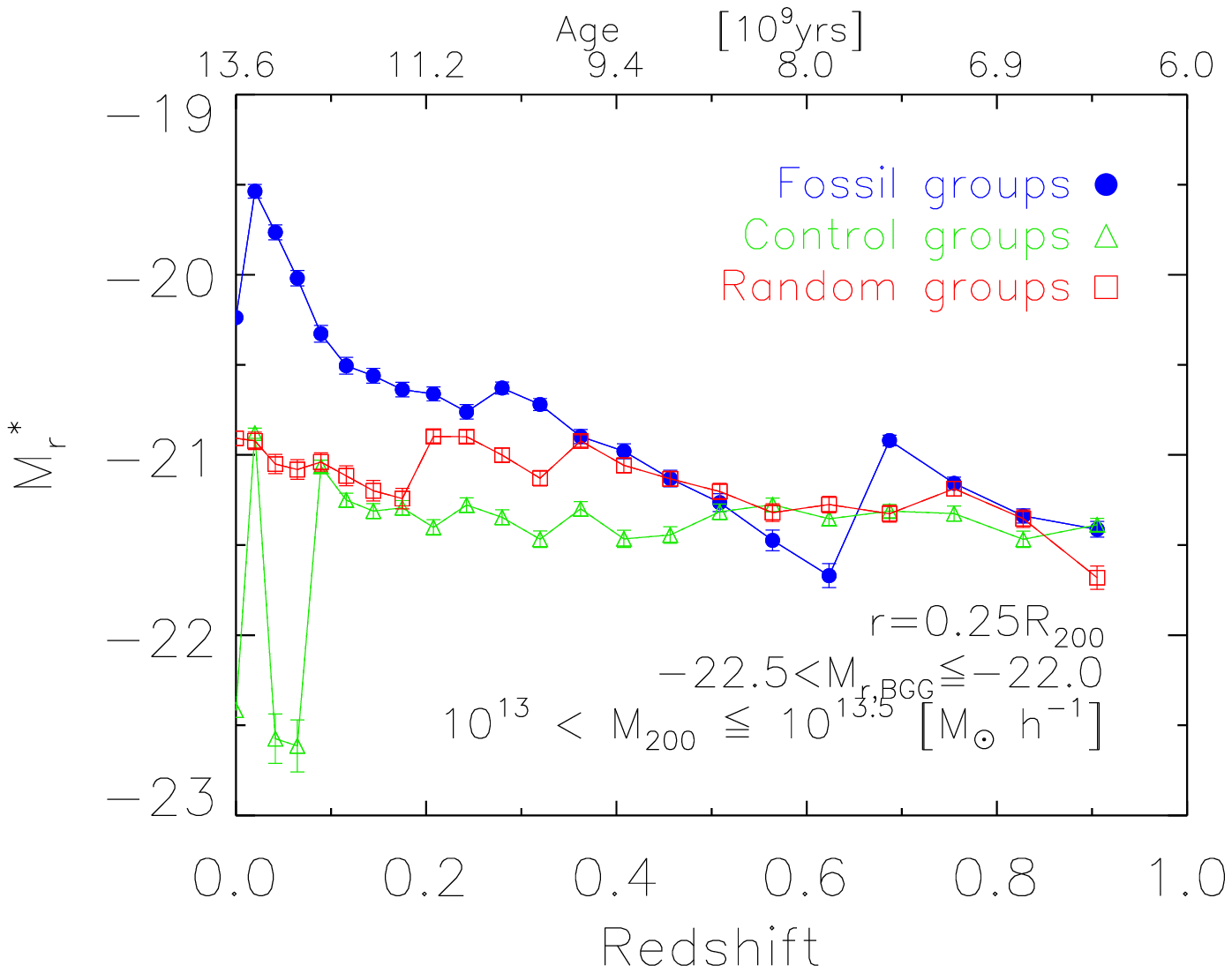}
   \includegraphics[width=8.5cm]{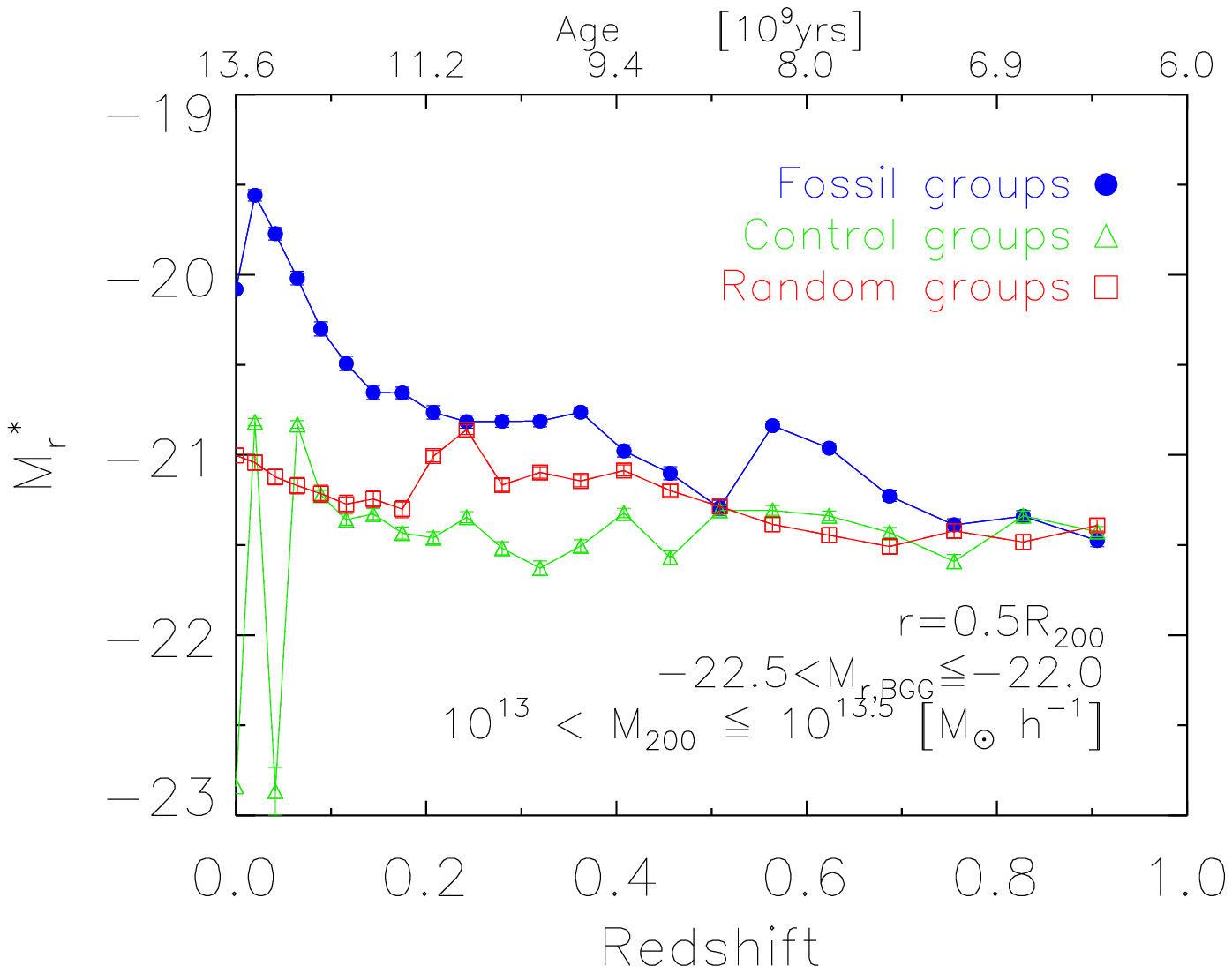}
   \includegraphics[width=8.5cm]{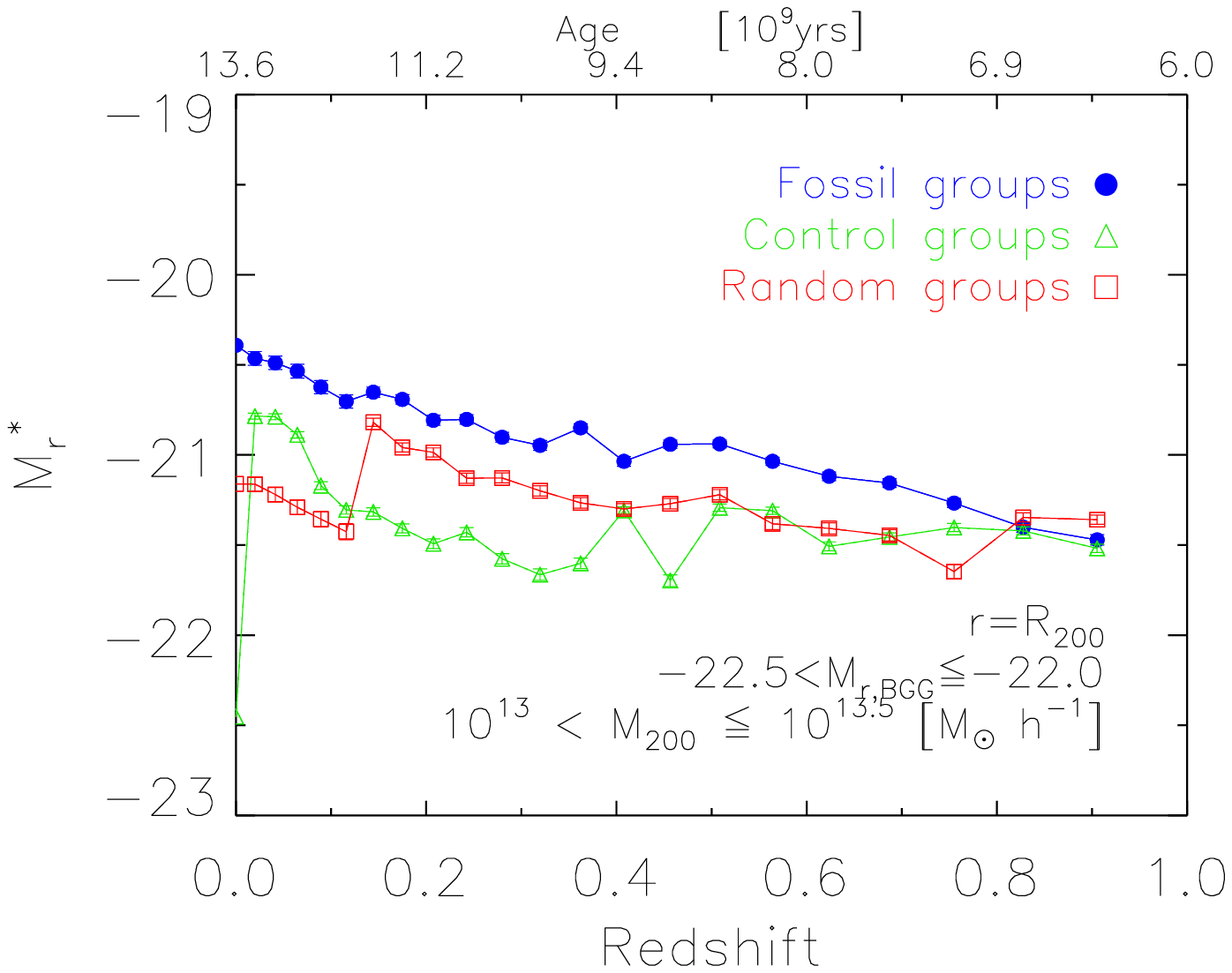}
  
        \end{center}
        \caption[flag]{Evolution of $ M^{*}$ with redshift and the age
          of the universe for fossil (blue filled circles), control (green
          open triangles), and random groups (red open squares) for
          SI. Panels from top to bottom present $ M^{*}$ evolution of
          the composite LF of galaxies which computed within
          $\frac{1}{4} R_{200}$, $ \frac{1}{2} R_{200}$, and $R_{200}$,
          respectively. }
      \label{mstar_m1}
 \end{figure} 

  \begin{figure}
    \begin{center}  
      \leavevmode
    \includegraphics[width=8.5cm]{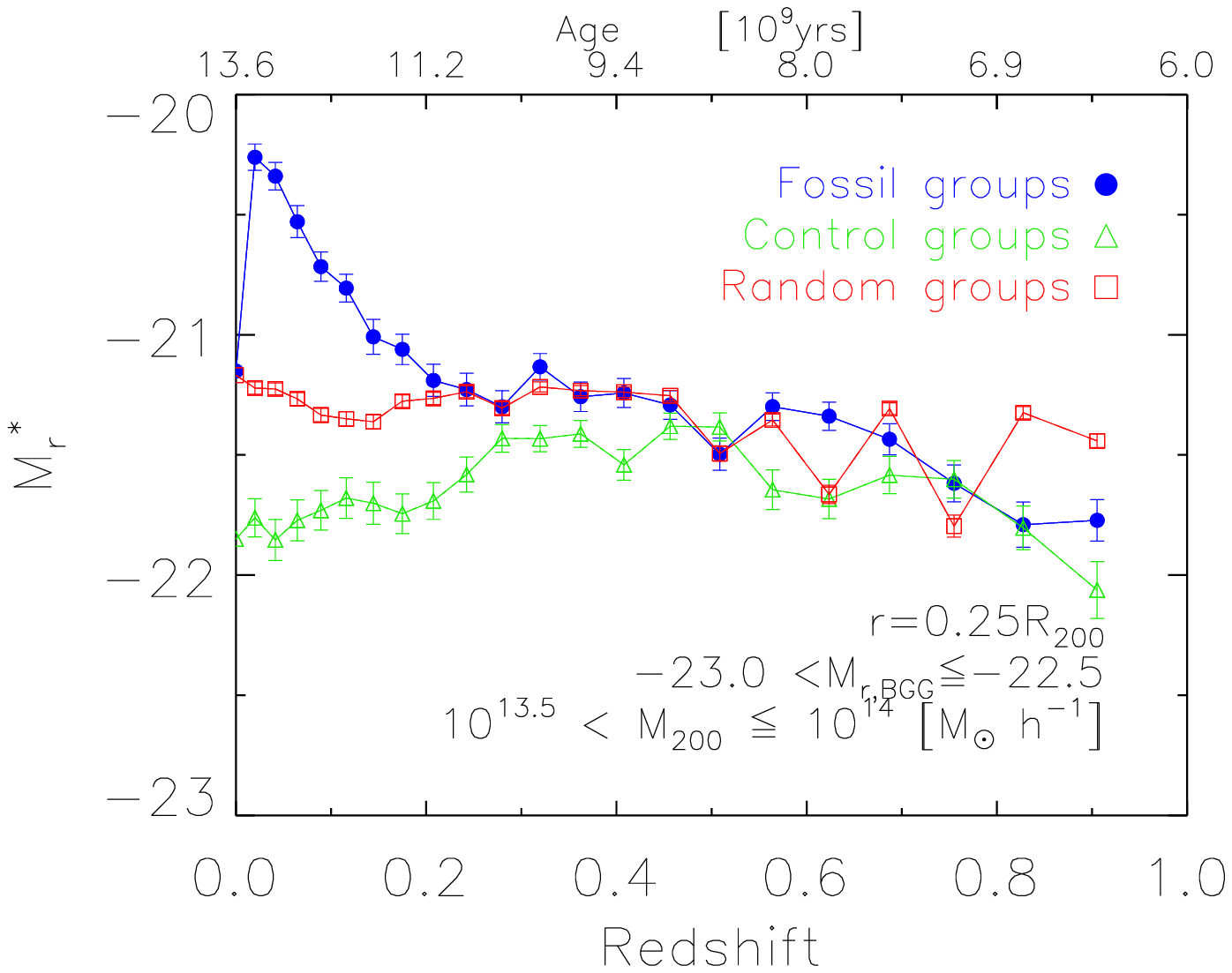}
     \includegraphics[width=8.5cm]{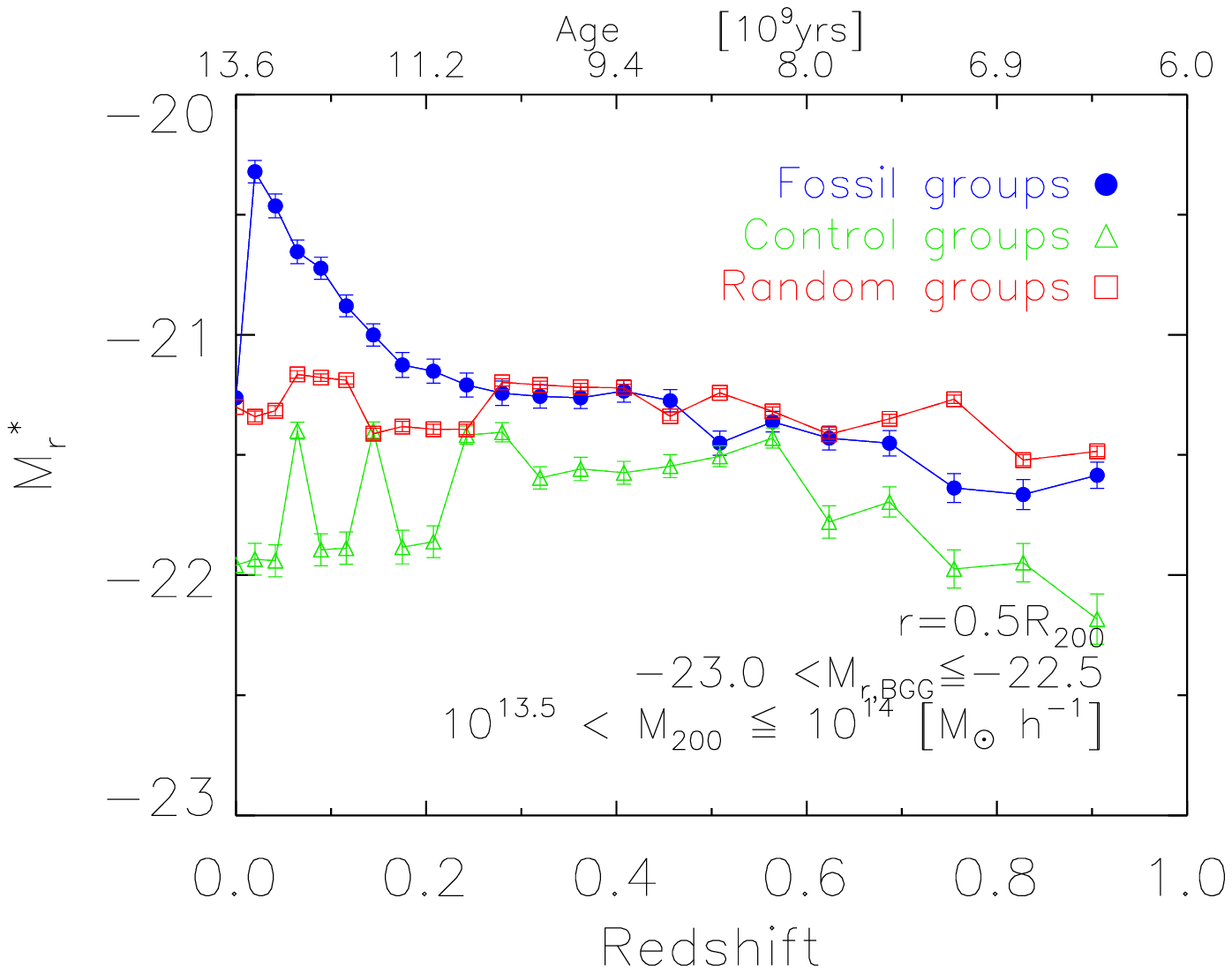}
     \includegraphics[width=8.5cm]{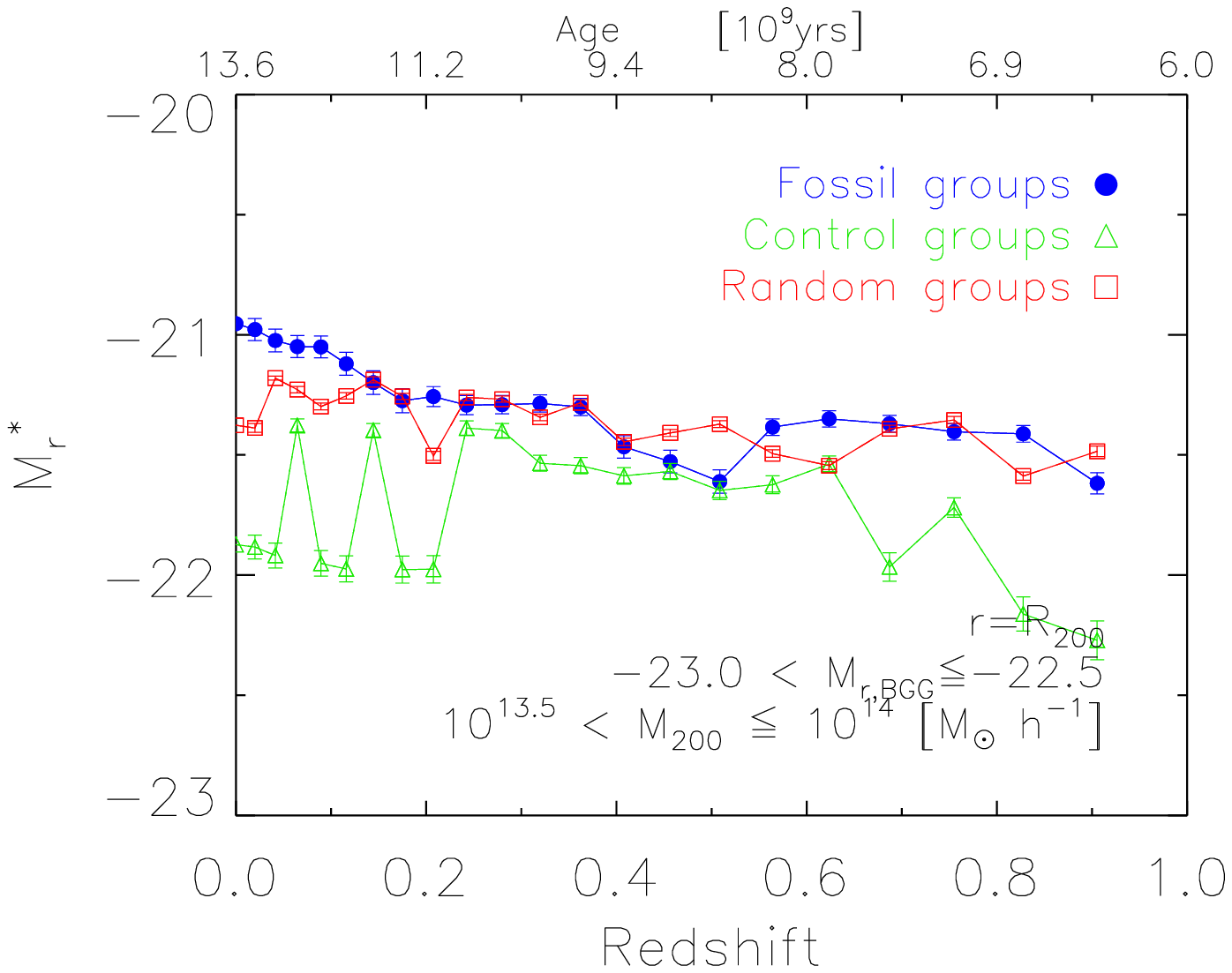}
   
         \end{center}
         \caption[flag]{As in Fig. \ref{mstar_m1}, except for SII. }
       \label{mstar_m2}
  \end{figure}
    
If fossil groups are indeed early-formed systems, it is feasible to
address this by studying the redshift evolution of the two particular
components of galaxy LF, bright, and faint ends, with the help of
SAMs. To this end, we select our galaxy group samples at z=0 according
to the definition given in \S 2.2. Then, we trace backwards all halos
and measure the composite LF of galaxies in groups/clusters for each
sub-sample in 22 different redshift slices from $ z=0 $ to $ z=1 $, down to
an absolute r-band magnitude of -15 mag. The composite LF is
measured from a combination of the LF of traced galaxy groups, 
according to the method introduced in detail by
\cite{Colles89}. The number of galaxies within each luminosity or
magnitude bin is measured using the following equation (1):
 
  \begin{equation}
 N_{cj}=\frac{N_{c0}}{m_{j}}\sum_{i} \frac{N_{ij}}{N_{i0}}   \hspace{6mm}   
  \end{equation}
  where $N_{cj}$ is number of galaxies within the j-th bin of the
  composite LF, $N_{ij}$ is the count in the $j$-th bin of the $i$-th
  group$^{,}s$ LF, $N_{i0}$ is the normalisation of the $i$-th group
  LF (assumed here the number of galaxies brighter than $M_{r} =-19$
  within given radius), $m_j$ is the number of groups counterpart to
  the $ j$-th bin, and
  
  \begin{equation}
  N_{c0}=\sum_{i}N_{i0}   \hspace{6mm}      
  \end{equation}
  
  The error for each bin of the composite LF is calculated
  according to the following relation:
 
 \begin{equation}
\delta N_{cj}=\frac{N_{c0}}{m_{j}}[\sum_{i} (\frac{\delta N_{ij}
}{N_{i0}})^{2}]^{1/2}   \hspace{6mm}   
\end{equation}
where $ \delta N_{ij} $ and $ \delta N_{cj} $ and are formal errors
corresponding to the $j$-th bin of the luminosity function of the
$i$-th galaxy groups and composite groups, respectively.
 
In order to study the environmental dependence of the composite LF, we
compute the LF of each group within three radii from the BGG:
$\frac{1}{4}R_{200},\frac{1}{2}R_{200}$, and$ R_{200}$. To quantify the
redshift evolution of the composite LF of galaxies in fossil, control, 
and random groups, we fit a single Schechter function
\citep{Schechter06}:
   
   \begin{equation}
 \phi(M)dM = (0.4 ln10)\phi^{*}X^{(1+\alpha)}\times exp^{-X}dM   
 \end{equation}
 where $X=10^{0.4(M^{*}-M )}$ and $\phi(M)$ is proportional to the
 number of galaxies that have absolute magnitudes in the range (M,
 M+dM). The  $\phi^{*}$ parameter is the characteristic number of galaxies and $M^{*}$
 is the characteristic absolute magnitude. The
 Schechter function drops sharply at bright magnitudes and rises at the faint end following
 a power law with a slope given by $\alpha$.
    
 The large magnitude gap in the LF of fossil groups, affects the
 goodness of the fit. To overcome this effect, we exclude the
 BGGs/BCGs from the Schechter function fit to the composite LF.

 Fig. \ref{LFs} shows the Schechter function fit (green dashed curve)
 to the composite LF of galaxies (black histogram with error bars) in
 fossil groups/clusters within $\frac{1}{4}R_{200}$ (top panel),
 $\frac{1}{2}R_{200}$ (middle panel), and $ R_{200}$ (bottom panel)
  for SIII at $ z=0$. As discussed in \S 2.2, applying limits on
 BGG/BCG luminosities in defining group/cluster sub-samples, allows us
 to detect the large magnitude gap  in the bright end of the composite LF of galaxies in
 fossil groups (see Fig. \ref{LFs}).  In Fig.\ref{sigma}, we show
 the 1-$ \sigma $ and $ 2-\sigma $ error contours of the fit on the
 composite LF fossils at $ z=0 $.

 \begin{figure}
   \begin{center}  
     \leavevmode
  \includegraphics[width=8.5cm]{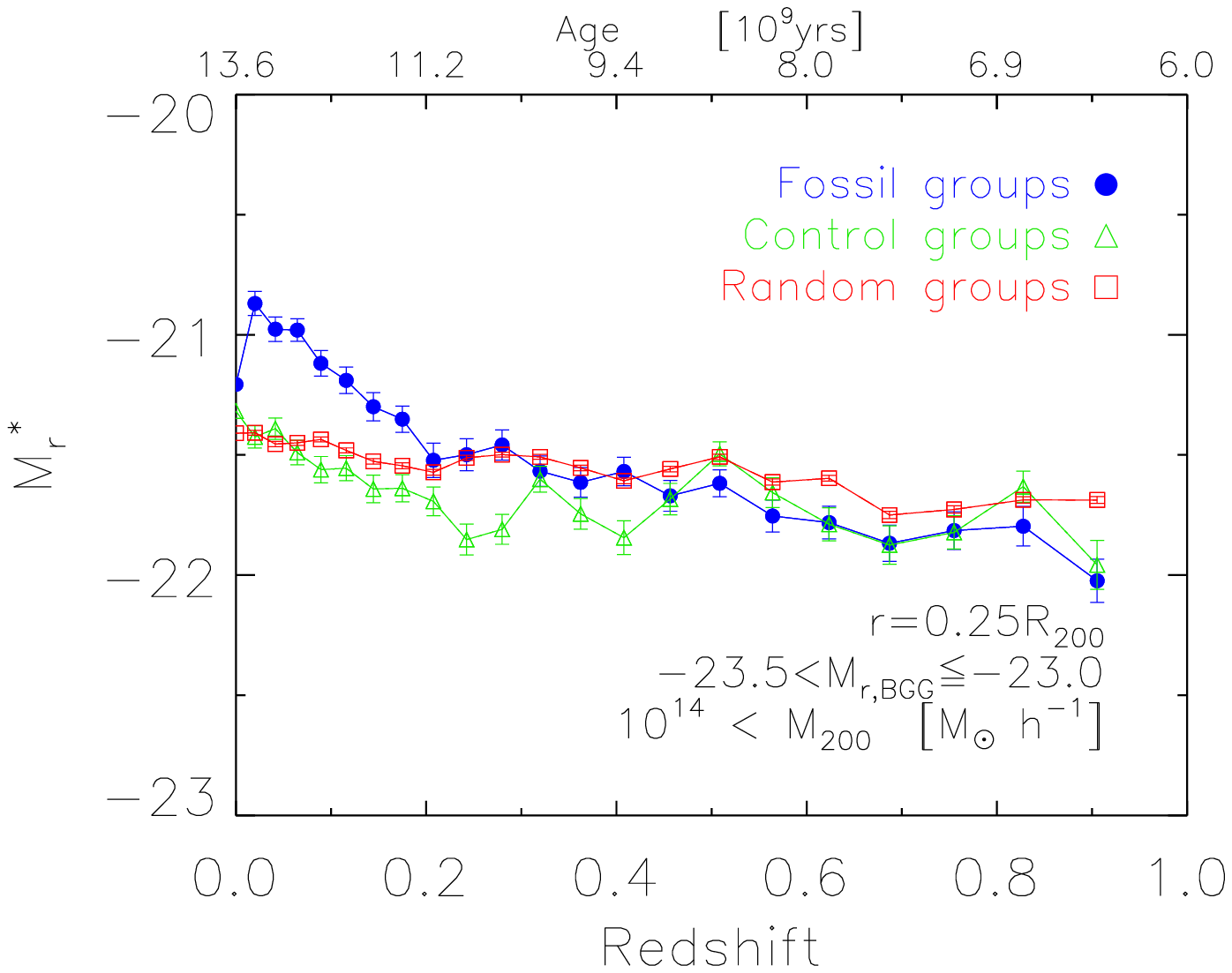}
   \includegraphics[width=8.5cm]{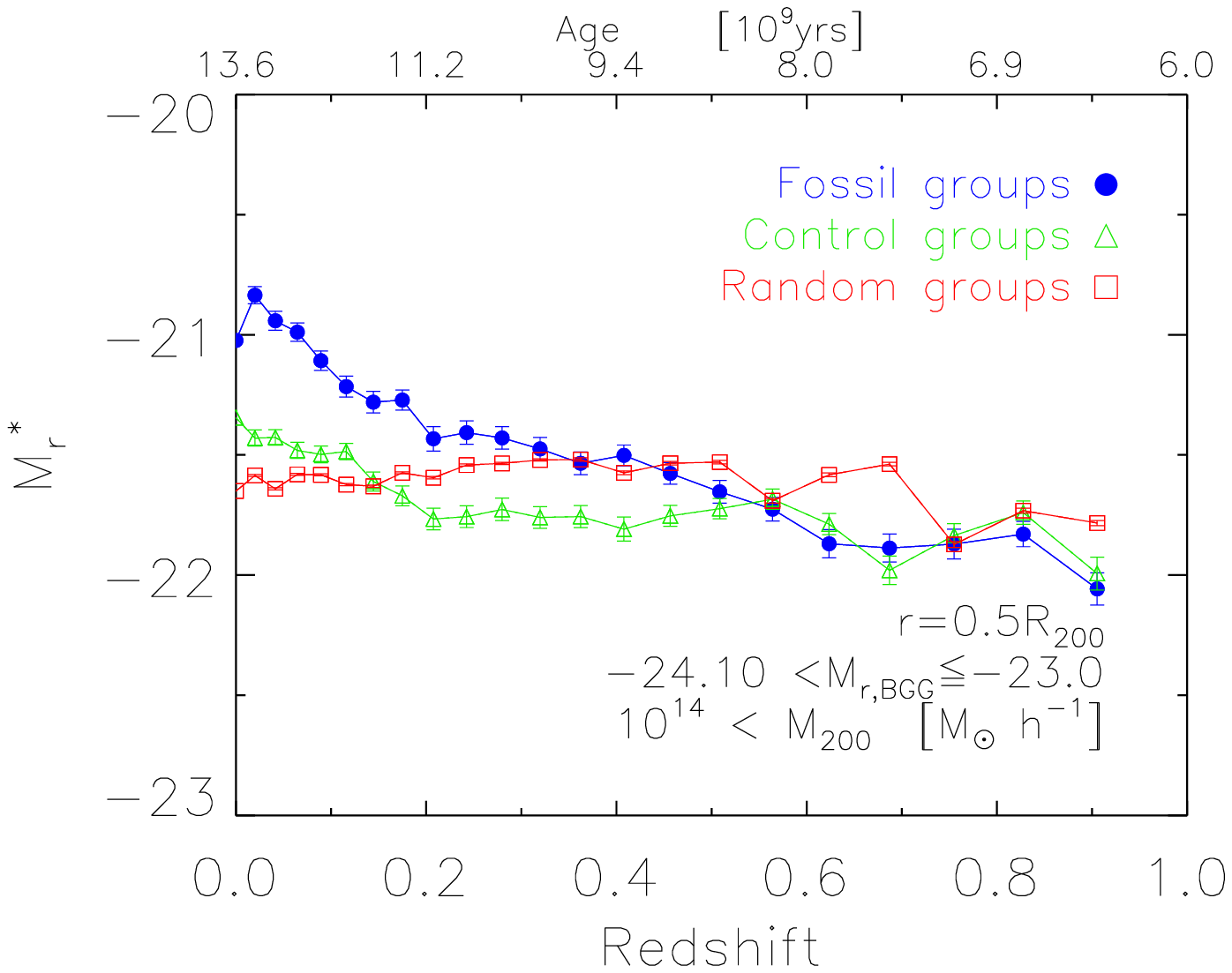}
   \includegraphics[width=8.5cm]{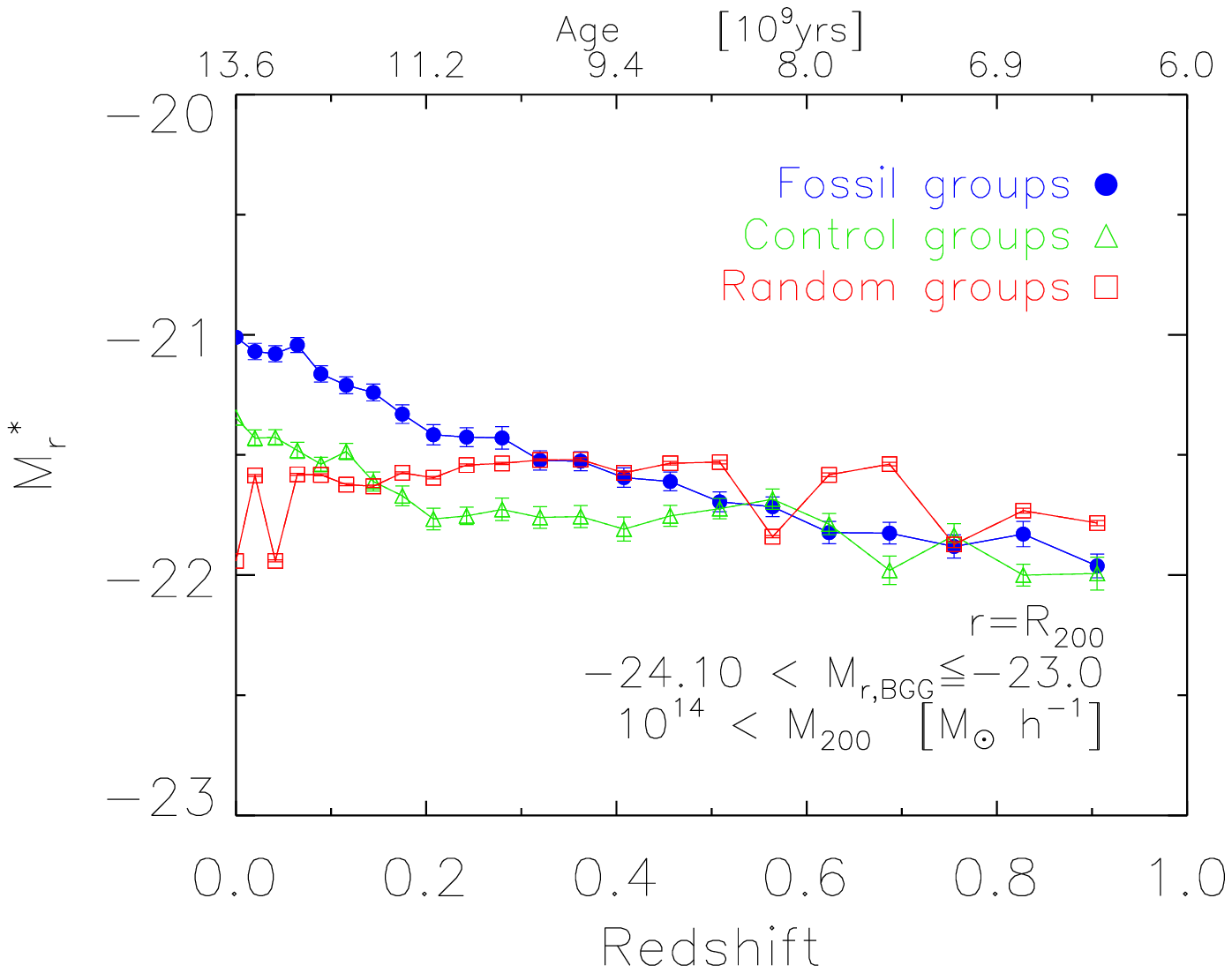}
  
        \end{center}
        \caption[flag]{As in Fig. \ref{mstar_m1}, except for SIII.  }
      \label{mstar_m3}
 \end{figure}
  
\subsection{Evolution of the luminosity of the bright end of LF, $ M^{*} $ }

In Fig. \ref{mstar_m1}, we plot the $ M^{*}$ versus redshift for
fossils (filled blue circle) , controls (green open triangle), and
random groups (red open square) for SI.  The luminosity of the bright end
of the composite LF of galaxies in fossil groups, selected within
$\frac{1}{4} R_{200} $ radius from BGG evolves substantially, changing
$M^*$ by $ \sim2 $ mag, while LF of random groups shows no
considerable evolution with redshift. The bright end of the composite
LF of the control groups shows no evolution between z$\sim0.1 $ and
$z= 1 $. However, it appears that at $z<0.1 $, $ M^{*}$ of these
groups is skewed to lower magnitudes, which is probably caused by the
infall of luminous galaxies within $\frac{1}{4}R_{200} $.

As shown in the middle panel of Fig. \ref{mstar_m1}, we find 
evolution of $ M^{*}$ for the composite LF of galaxies selected within
$ \frac{1}{2} R_{200} $ for different types of groups for SI are similar to that of the composite LF within $ \frac{1}{4}R_{200} $. We
notice that by increasing the group radius to $ R_{200}$, $ M^{*}$ of
fossils increases between $ z=1 $ and present day by $ \sim 1 $ mag (see
lower panel in Fig. \ref{mstar_m1}).  As a result, the $ M^{*}$ of the
composite luminosity function of galaxies in fossil groups within $
R_{200} $ at $ z\sim0.05 $ is consistent with that ($ M^{*}=21.10$) of
our study of the composite luminosity function of four fossil group
candidates \citep{Khosroshahi14}. No significant redshift evolution of
$ M^{*}$ is seen for the random groups and control groups.

Our conclusions derived for SI sub-sample also apply to SII and SIII
(see Fig.\ref{mstar_m2} and Fig.\ref{mstar_m3}).  For both SII and
SIII sub-samples, the $ M^{*}$ of composite LF of galaxies within $
\frac{1}{4} R_{200} $ and $\frac{1}{2} R_{200} $ radii of fossil
groups shows an evolution with redshift by $ \approx 1-1.5 $ mag,
which is different from $ M^{*}$ evolution of LF for the random and
control groups. We notice a substantial change in $ M^{*}$ of
composite LF of fossil groups takes place around the central region of
these systems at z$<0.4 $ in the last $ \sim 5 Gyr $. Furthermore, the
slope of $ M^{*}-z$ relation of fossil groups decreases with 
increasing halo mass. 

In this study, we have selected our groups by considering the r-band
absolute magnitude of BGGs, magnitude gap- and halo mass of
groups/clusters. These considerations do not allow us to compare our
findings with previous results from other literature. Nevertheless,
our prediction of the $ M^{*}/\alpha$ evolution  is
approximately consistent with that of the $ r^{\prime}$-band composite
LF of galaxies in C1 clusters presented in Table 4 in
\cite{Alshino10}. We perform a comprehensive comparison of the
evolution of the galaxy  composite LF in groups/clusters from the
SAM with that of observations using the X-ray galaxy groups catalogue
in Gozaliasl et al. (in prep.).

 \begin{figure}
   \begin{center}  
     \leavevmode
   \includegraphics[width=8.5cm]{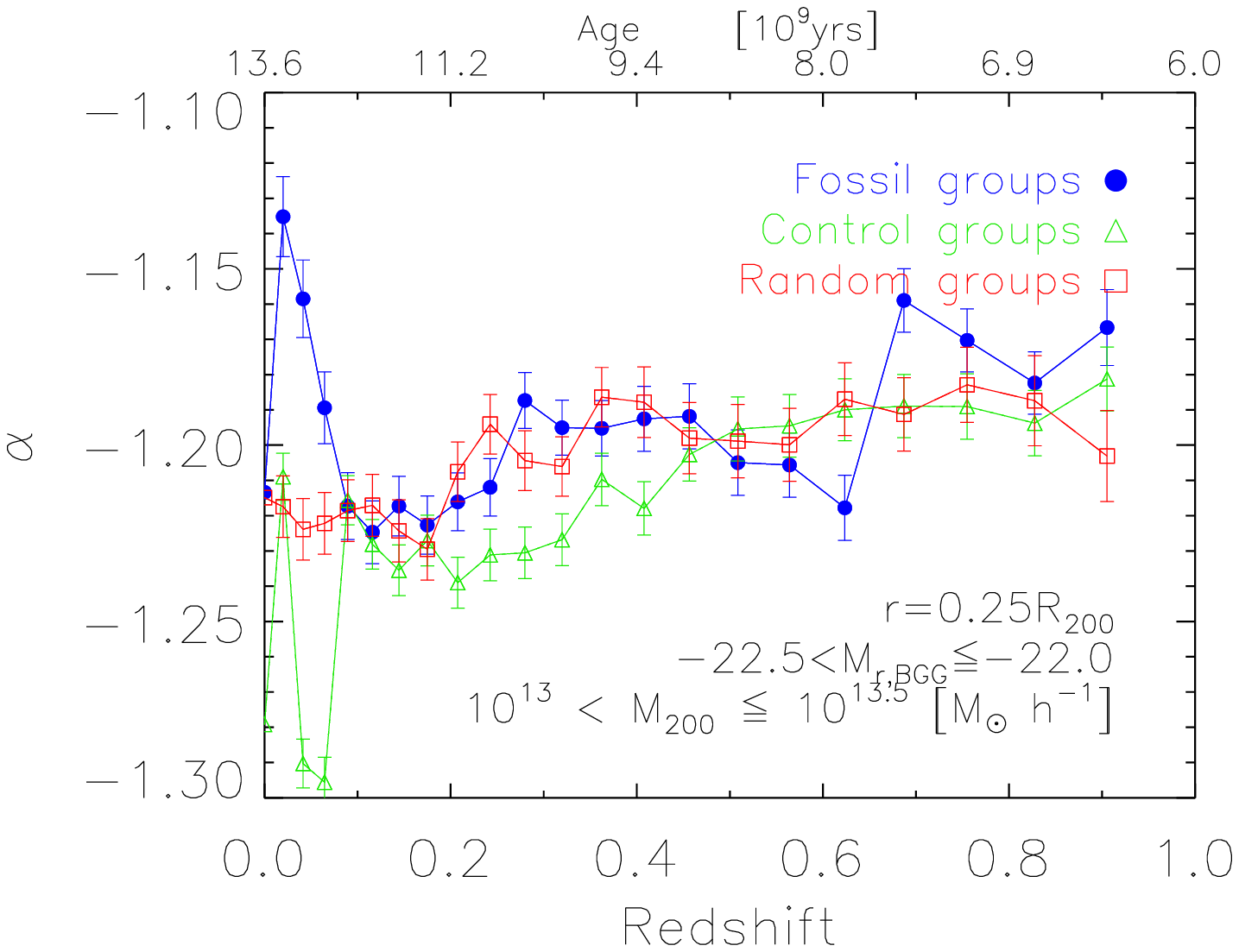}
   \includegraphics[width=8.5cm]{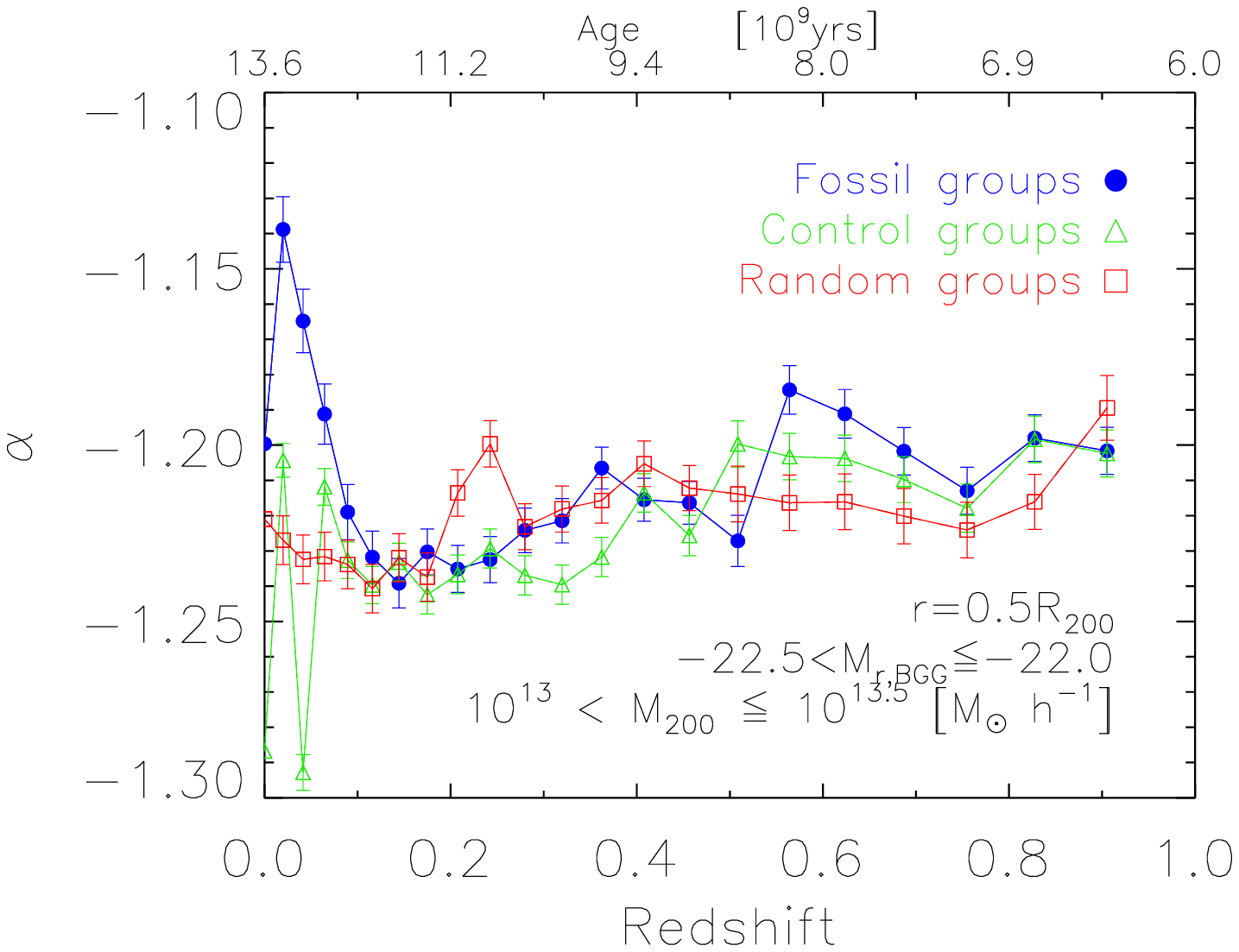}
   \includegraphics[width=8.5cm]{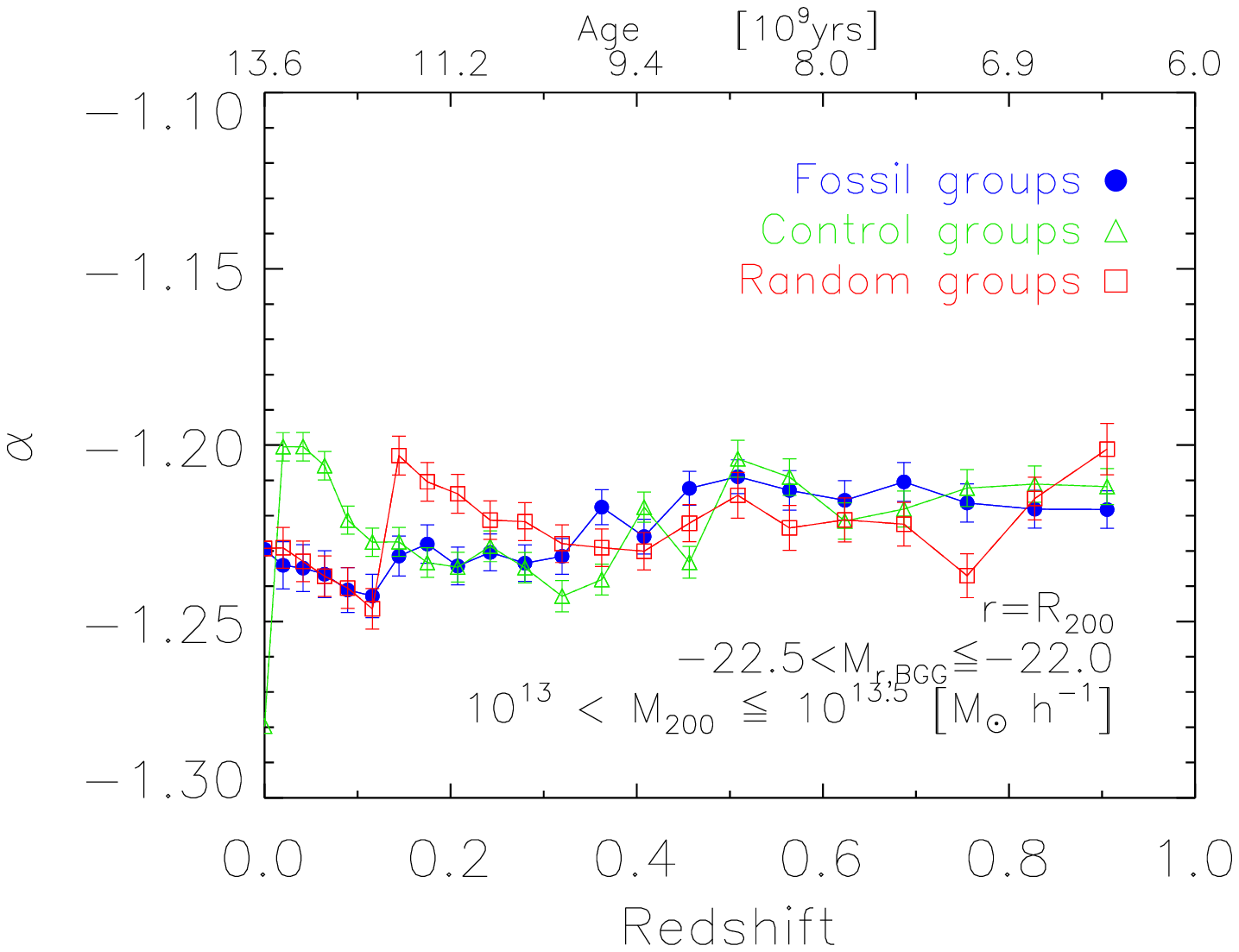}
  
        \end{center}
        \caption[]{Evolution of $ \alpha$ with redshift and the age of
          the universe for fossil (blue filled circles), control (green
          open triangles), and random groups (red open squares) for
          SI. Panels from top to bottom present evolution of the faint
          end slope, $\alpha $, of the composite LF of galaxies within
          $\frac{1}{4} R_{200}$, $ \frac{1}{2} R_{200}$, and $R_{200}$,
          respectively. }
      \label{alpha1}
 \end{figure}
 
 \begin{figure}
   \begin{center}  
     \leavevmode
   \includegraphics[width=8.5cm]{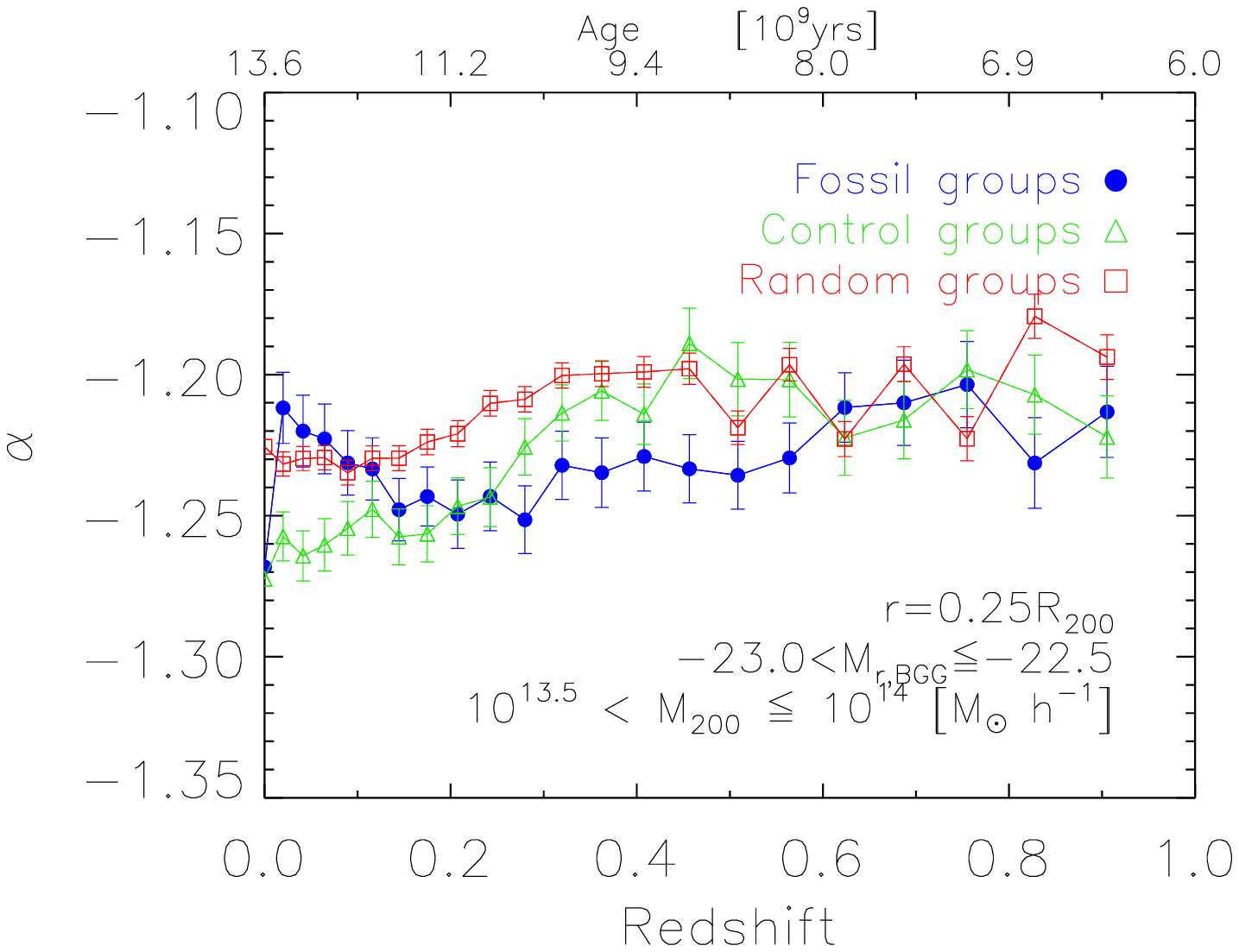}
    \includegraphics[width=8.5cm]{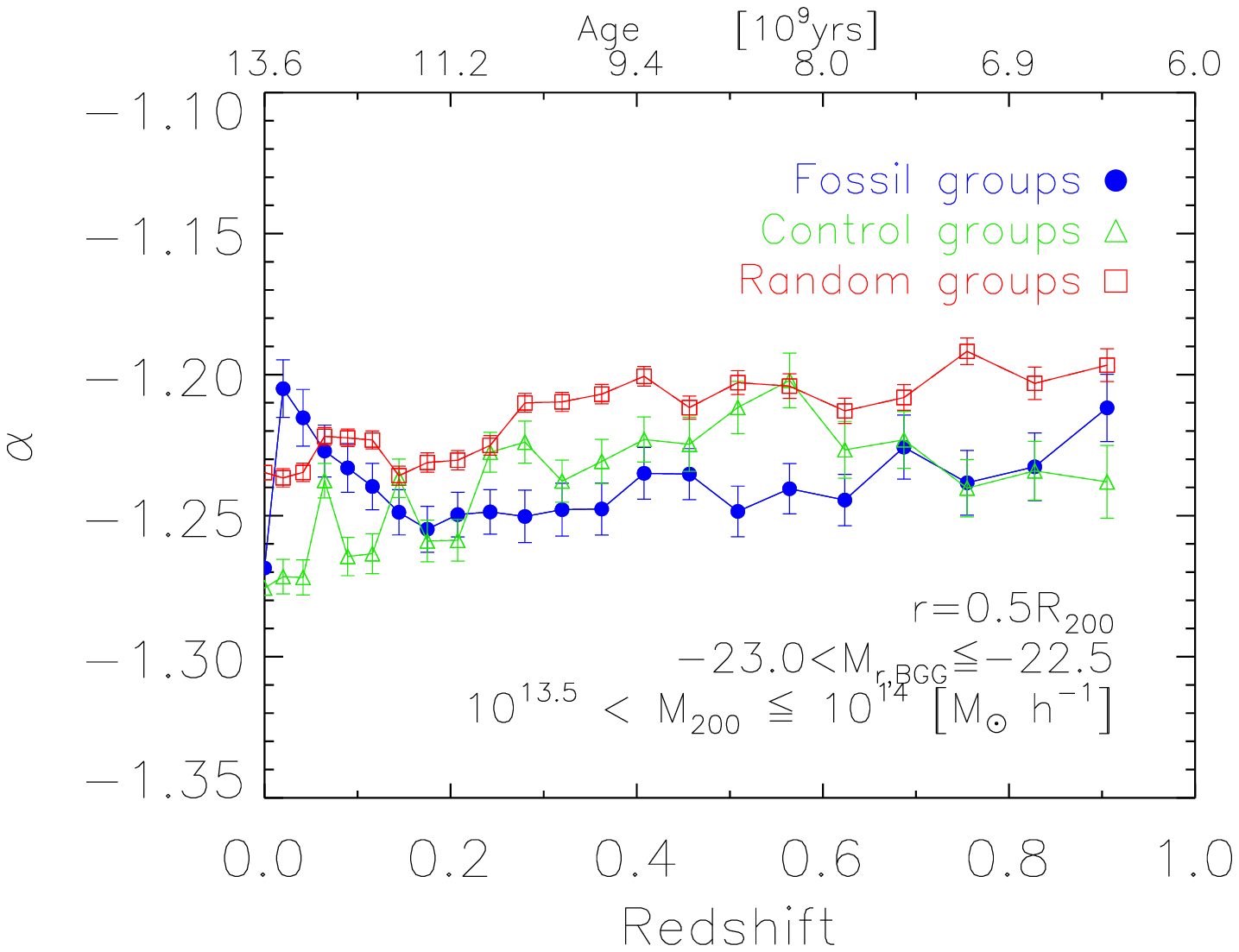}
    \includegraphics[width=8.5cm]{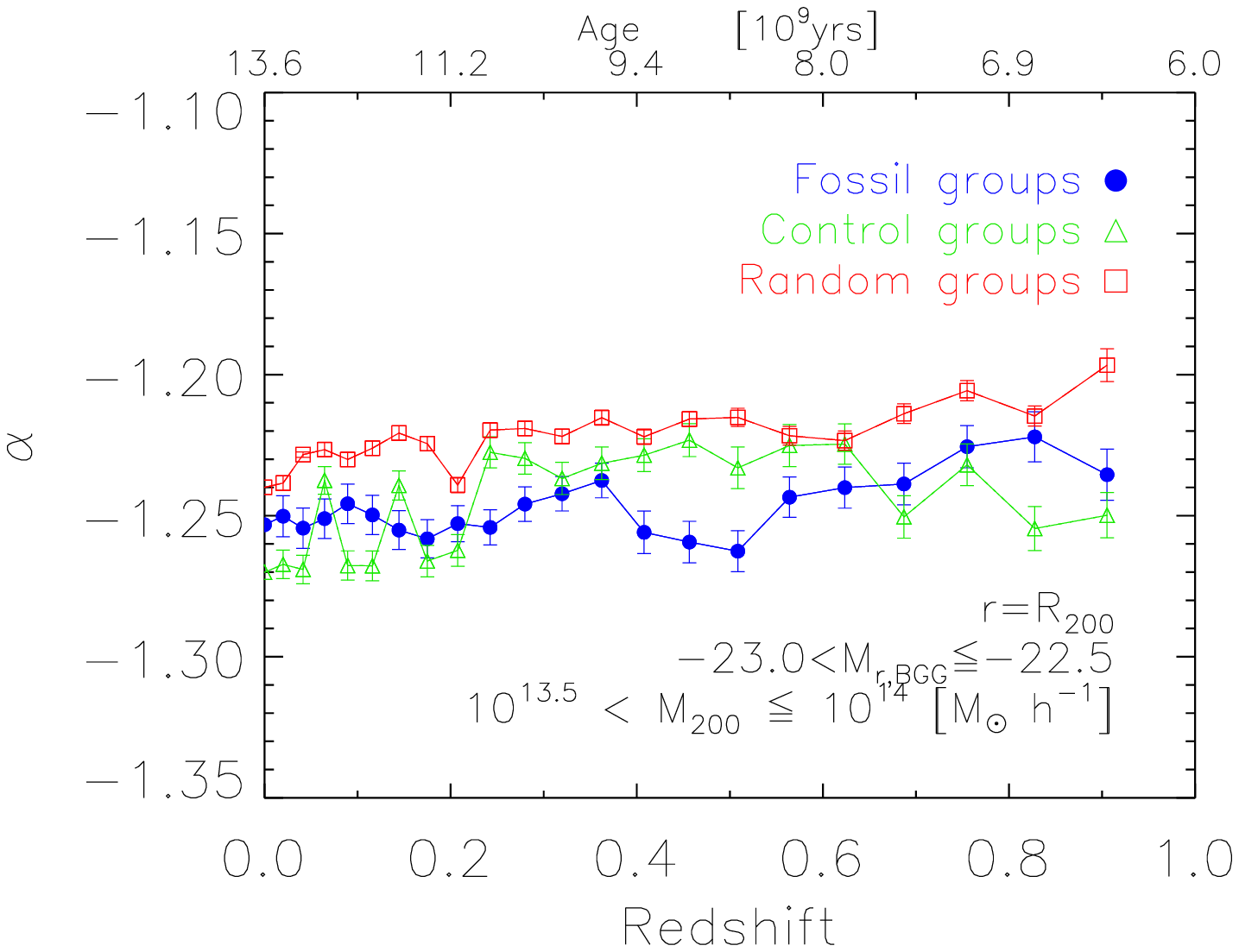}
  
        \end{center}
        \caption[flag]{As in \ref{alpha1}, except for SII.}
      \label{alpha2}
 \end{figure}

 \begin{figure}
   \begin{center}  
     \leavevmode
  \includegraphics[width=8.5cm]{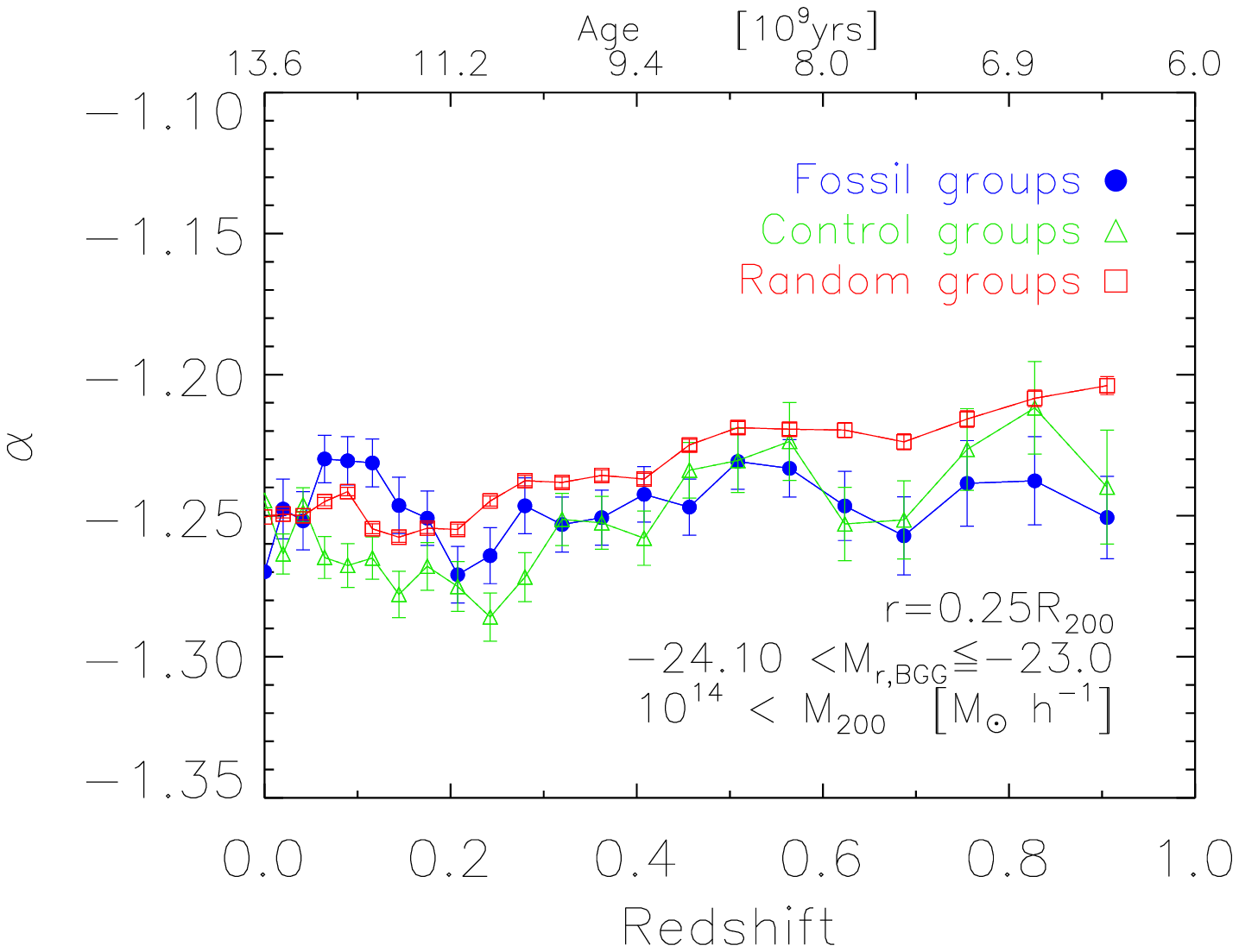}
   \includegraphics[width=8.5cm]{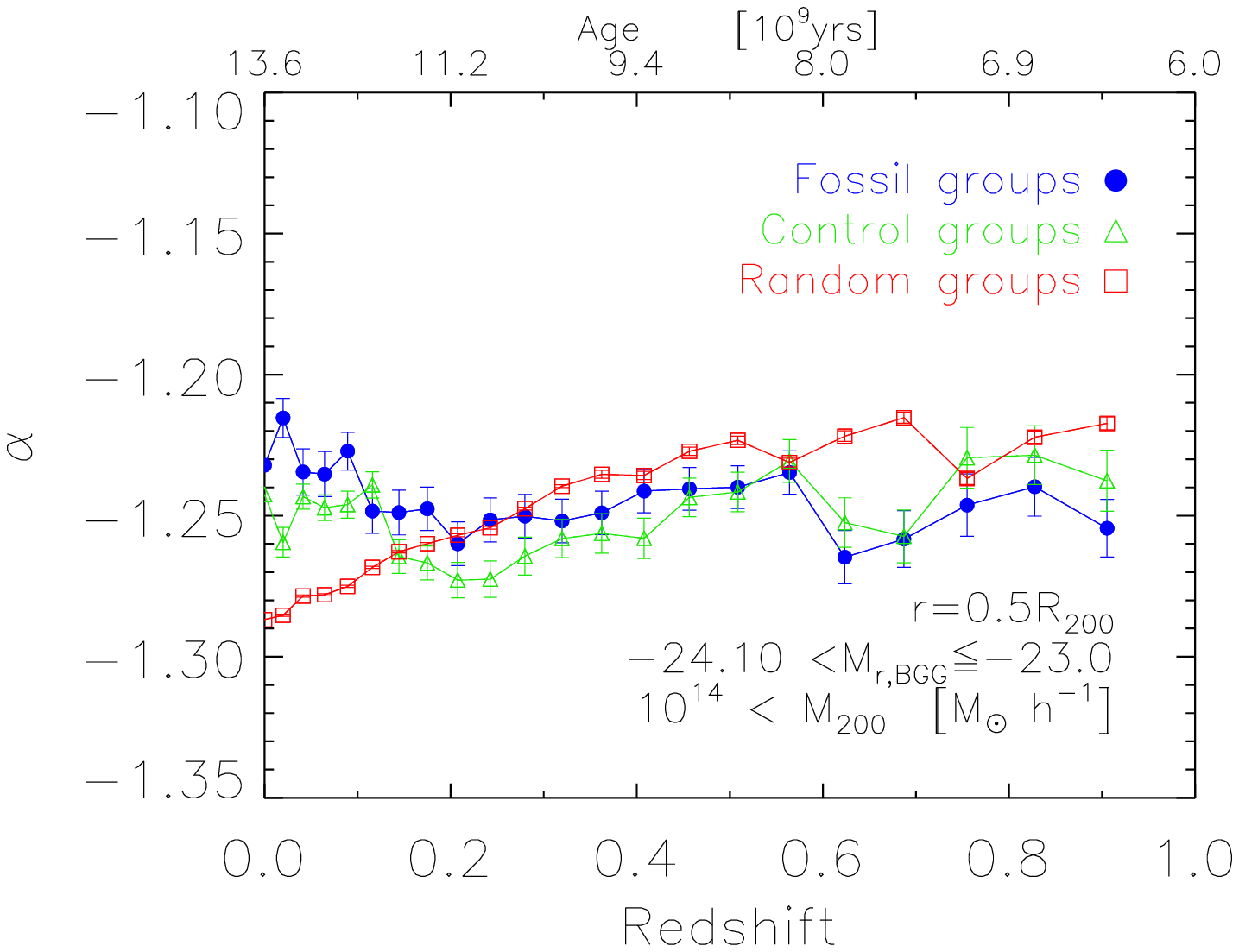}
   \includegraphics[width=8.5cm]{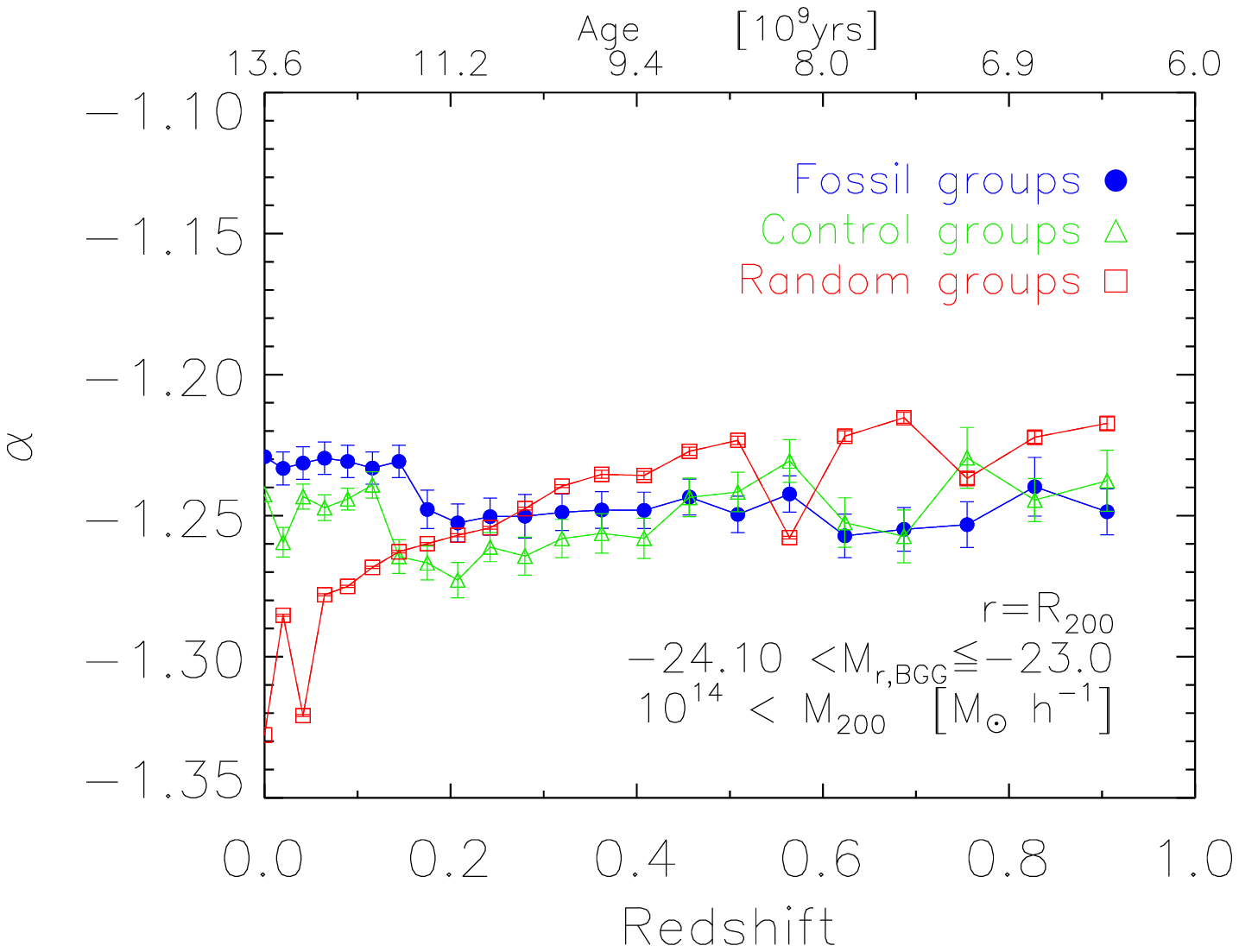}
  
        \end{center}
        \caption[flag]{As in \ref{alpha1}, except for SIII. }
      \label{alpha3}
 \end{figure}

\subsection{Evolution of the faint end slope of LF, $ \alpha $ }

In Fig. \ref{alpha1}, we show the redshift evolution of the faint end
slope, $ \alpha $, of the composite LF of galaxies corresponding to
$\frac{1}{4} R_{200}$ (top panel), $ \frac{1}{2} R_{200}$ (middle
panel), and $R_{200}$ (bottom panel) for SI.  The faint end slope of
composite LF of galaxies within $\frac{1}{4} R_{200}$ for fossil,
control, and random groups show little evolution with redshift
between z$=1 $ and z$=0.1$. At z$<0.1 $ the slope starts evolving
steeply with redshift from $ \alpha\sim-1.22 $ to $ \alpha\sim-1.14$
for the fossils. It appears that the faint end of composite LF of
control groups evolves from $ \alpha\sim-1.22 $ to $
\alpha\sim-1.3$. The $ \alpha $ of the composite LF of random groups
shows no evolution in this redshift range. A similar evolution of $
\alpha $ is also detected for the composite LF of galaxies within
$\frac{1}{2} R_{200}$ (middle panel of Fig. \ref{alpha1}). The faint
end slope of the composite LF of galaxies in all three types of groups
within $ R_{200}$ shows no evolution with redshift. In addition, the
predicted $\alpha $ of the composite luminosity function of fossil
groups within $ R_{200} $ at $ z\sim0.05 $ agrees within observational
uncertainties with that ($\alpha=-1.32$) of our study of the composite
luminosity function of four fossil group candidates
\citep{Khosroshahi14}.

In Fig. \ref{alpha2}, we illustrate $ \alpha $ as a function of time
and redshift for SII. All three panels indicate $ \alpha $ has
approximately similar constant behaviour with redshift. The $ \alpha $
of composite LF of fossil groups (for SI and SII) at
$\lesssim\frac{1}{2} R_{200}$ shows an evolution at $ z\lesssim0.2 $
by $\sim0.05-0.1$.  This evolution of $ \alpha $ indicates that the
number of the dwarf galaxies within central regions of fossil groups
at $\lesssim\frac{1}{2} R_{200}$ reduces with time, which can be a
consequence of the merger ratio within these systems.

Fig. \ref{alpha3} presents a redshift evolution of the slope of the
faint end of the composite LF for SIII.  We observe no significant
redshift evolution of $ \alpha $ ($\sim$0.05) for all types of selected
groups within $\frac{1}{4} R_{200}$, $\frac{1}{2} R_{200}$, and $
R_{200}$.

\begin{figure}
  \begin{center}  
    \leavevmode
 \includegraphics[width=8.5cm]{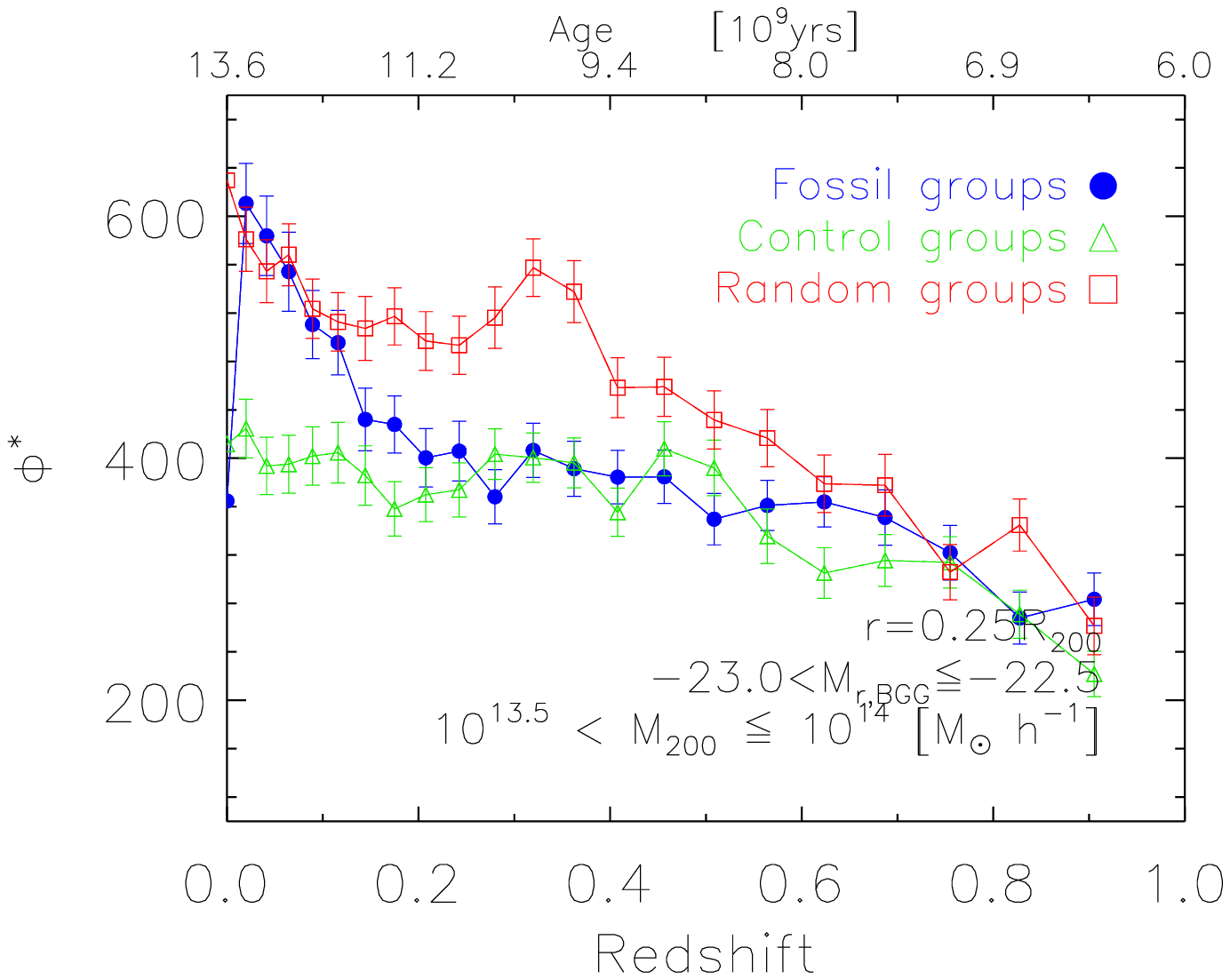}
  \includegraphics[width=8.5cm]{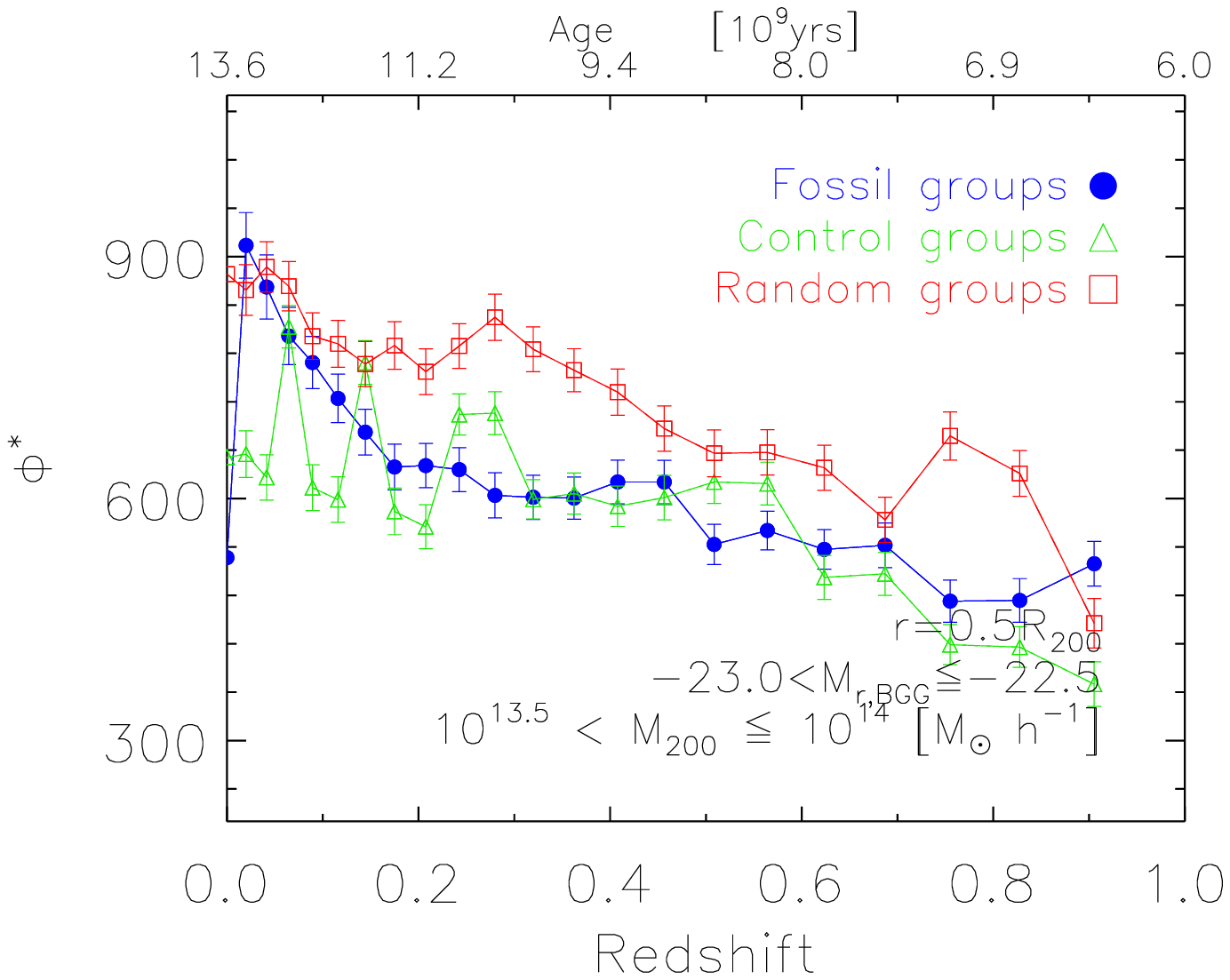}
  \includegraphics[width=8.5cm]{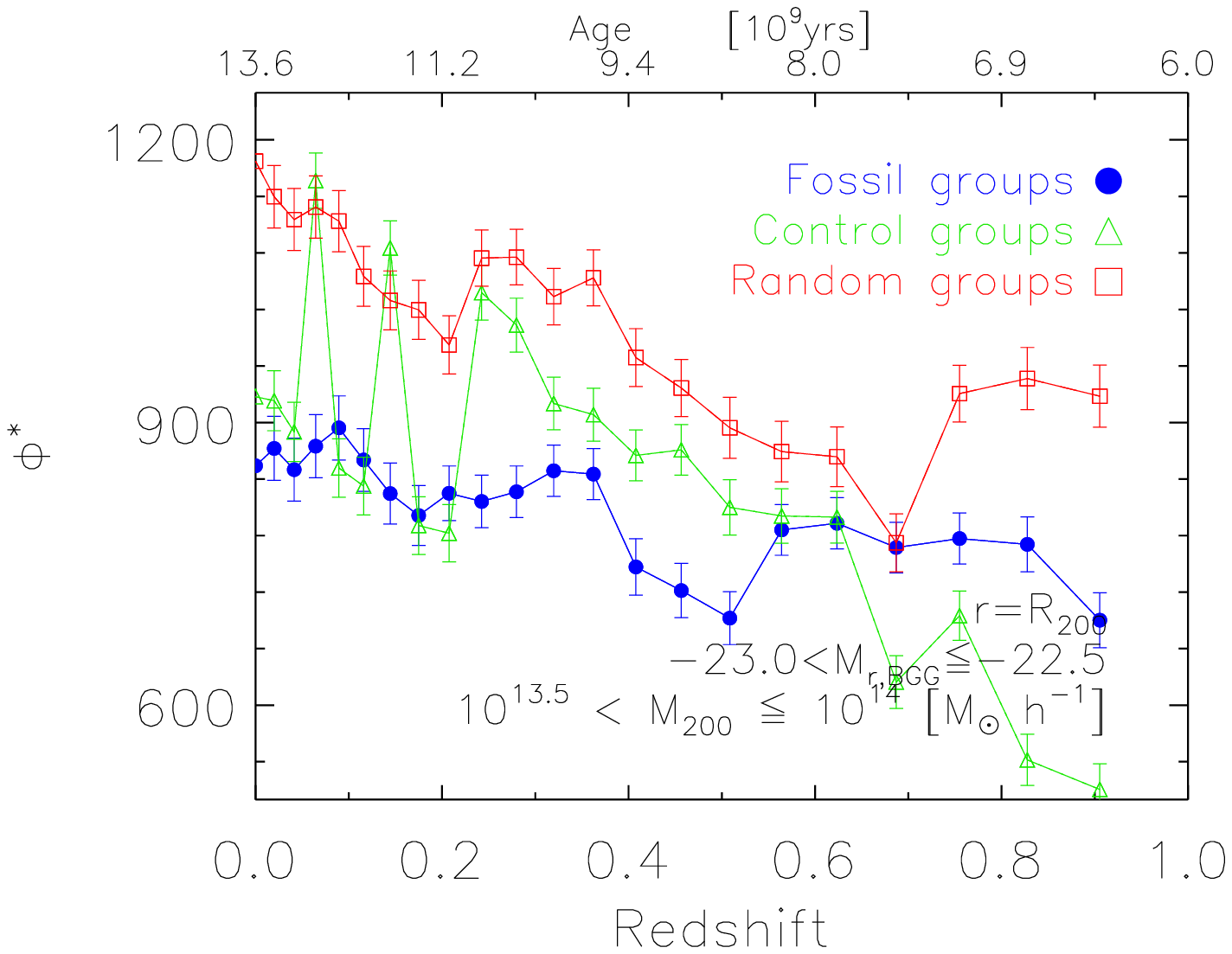}
 
       \end{center}
       \caption[flag]{Evolution of $ \phi^{*}$  with redshift and the age of the universe for fossil (blue points), control (green point), and random (red point) groups for SII. Panels show this evolution for composite LFs for $\frac{1}{4} R_{200}$ (top panel), $\frac{1}{2} R_{200}$ (middle panel), and $R_{200}$ (bottom panel). A similar trends are detected for the evolution of $ \phi^{*}$ for groups within  SI and SIII.}
     \label{phi_m1}
\end{figure}

 \subsection{Evolution of  $ \phi^{*} $ and dwarf galaxies}

 In Fig. \ref{phi_m1}, we show $ \phi^{*} $ of the composite LFs of
 galaxy groups versus redshift for SII. The $ \phi^{*} $ of the
 composite LFs of groups in SI and SIII are similar to that of
 SII. Each panel corresponds to the $ \phi^{*} $ evolution of the
 composite LFs of groups within $\frac{1}{4} R_{200}$ (upper panel), $
 \frac{1}{2} R_{200}$ (middle panel), and $R_{200} $ (lower panel). The
 $ \phi^{*} $ of the composite LFs of random and control groups show
 similar trends that increase with redshift in all panels. For fossil
 groups, we detect a trend for $ \phi^{*} $ of the composite LF
 within $ R_{200} $ (lower panel). We detect that $ \phi^{*} $ of the
 composite LFs of fossil groups within $\frac{1}{4} R_{200}$ and $
 \frac{1}{2} R_{200}$ also evolve similar to that of control groups
 between $ z\sim 1 $ to $ z\sim 0.25 $. However, the $ \phi^{*} $ of the composite LF of
 fossil groups rises rapidly from z$\sim0.2 $ to present day, indicating recent changes in the number of galaxies in these systems.

\begin{figure}
  \begin{center}  
    \leavevmode
 \includegraphics[width=8.5cm]{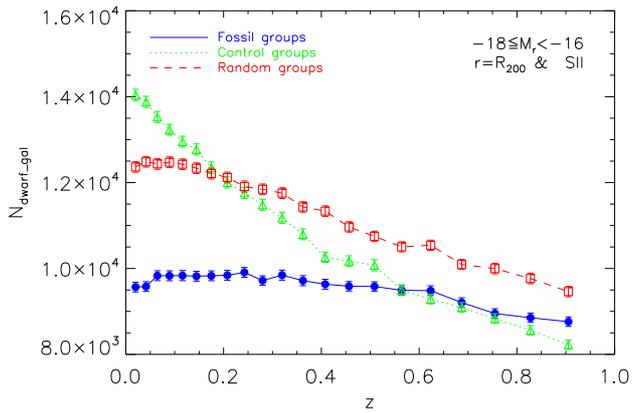}

       \end{center}
       \caption[flag]{Evolution of the number of dwarf galaxies with
         redshift for fossil (blue points), control (green point), and
         random (red point) groups for SII.  The $ N_{dwarf-gal}$ in
         fossil groups show no considerable evolution, while there is a
         significant evolution in control groups by $ \sim42\% $. The evolution of $ N_{dwarf-gal}$   for SI and SIII  are similar to that of SII. }
     \label{dwarf}
\end{figure}

In order to find out what caused the changes in $\phi^{*}$, we
investigate whether the number of dwarf galaxies ($ N_{dwarf-gal} $,
hereafter) in our sample of groups evolves with redshift. We
select galaxies in the r-band magnitude range $-18\leq M_{r} <
-16$. In Fig. \ref{dwarf}, we plot $ N_{dwarf-gal} $ of fossil (filled
blue circles), control (open green triangles), and random (open red
squares) groups within $ R_{200}$ versus redshift for SII (we find
similar trends for the $ N_{dwarf-gal}$ of groups in SI and SIII). The
$ N_{dwarf-gal}$ of fossil groups grows by $ \sim 12\% $ between $
z\sim1 $ and $z\sim0.3 $, after which $ N_{dwarf-gal}$ remains
constant with redshift. We also notice a $\sim 5$\% drop in $
N_{dwarf-gal}$ in fossil groups below $z=0.1$. This evolution is consistent with
the recent evolution of $ \alpha $ in composite LFs of fossils within
$ r<0.5R_{200} $ at $ z<0.2 $ (see the upper and middle panels of
Figs. \ref{alpha1} and \ref{alpha2}).

The $ N_{dwarf-gal}$ of random groups grows by $ \sim25\% $ since $
z\sim1 $ to present day.  We detect $N_{dwarf-gal}$ of control groups
to evolve similar to that of the fossils between $ z\sim1 $ to
$z\sim0.55$. In contrast to fossil groups, $ N_{dwarf-gal}$ of control
groups steeply rises with redshift by $ \sim45\% $ below $z\sim0.55 $.
The constant behavior of the number of dwarf galaxies in the fossil
groups indicates that changes in $ \phi^{*} $ are caused by changes in
$ M^{*} $ (see Figs. \ref{mstar_m1} to \ref{mstar_m3}) and not in the
number of galaxies. These findings strongly suggest that the BGG in
fossil group forms  because of the multiple mergers with massive galaxies,
and  that the large magnitude gap of fossil groups is the effect of galaxy
merger/cannibalism.

\section{Summary and conclusions}

In this paper, we investigate the redshift evolution of the composite
LF of galaxies within the mass-selected sub-samples of fossil groups,
control groups with $\Delta M_{1,2}\leq 0.5 $, and groups with any
$\Delta M_{1,2}$. We select our sub-samples at z=0 in the SAM of Guo et
al. (2011) and trace the halos backwards in redshift up to z$\sim 1
$. For 22 snapshots (redshifts), we calculate the mean luminosity gap,
and an average ratio of the halo mass of groups to their final masses
(at z=0). We also measure a composite LF of galaxies within three
radii, $\frac{1}{4} R_{200}$, $ \frac{1}{2} R_{200}$, and $
R_{200}$. We also trace forwards in time the magnitude gap of groups
from z=1 to present day. Our main results are as follows:

(I) Similar to other studies, we show that fossil groups in the G11 model
have accumulated their halo mass earlier than non-fossil groups. We
also detect that at redshift one the low-mass fossil systems (SI) had
on average $ \sim$15 per cent more mass compared to massive fossil
clusters (SIII).

(II) By tracing halos from $ z=0 $ to $ z=1 $, we show that the large
magnitude gaps of fossil groups for all halo mass ranges have been
formed between the present day and z$\sim 0.5$ (e.g. last $ \sim 5$
Gyrs). By forward tracing from $ z=1 $ to $ z=0 $ we find that 80 per cent
of fossil groups with $ M_{200} \leqslant 10^{14} $ M$ _{\sun} $h$
^{-1} $ fill their large gap possibly due to the infall of bright galaxies
within $ 0.5R_{200}$, in agreement with findings of
\cite{vonbendabeckmann08}. In addition, we detect 40 per cent of
fossil clusters with a halo mass above $10^{14} $ M$ _{\sun} $h$ ^{-1} $
retain their large magnitude gaps since $ z=1 $.

(III) We calculate composite LFs of galaxies in traced galaxy groups
at 22 redshift slices and fit a single Schechter function. The bright
end characteristic absolute magnitude, $ M^{*} $, of the composite LF
for fossils shows a significant redshift evolution by $ \sim -1.5$ mag
between z$\sim 0.5 $ and present day. This evolution for composite LFs
of fossils within $ 1/4R_{200} $ and $ 1/2R_{200} $ is comparable with
the composite LFs calculated inside $ R_{200} $. In contrast, the $
M^{*} $ of the composite LFs for control and random groups, corresponding to different radii, show no considerable evolution
 with redshift.

We observe that the slope of the faint end, $\alpha$, of the composite
LF within $ 1/4R_{200}$ and $ 1/2R_{200}$ for fossils with $
M_{200}\leqslant 10^{13.5} $ M$ _{\sun} $h$ ^{-1} $ evolves with
redshift by $ \sim 0.08 $ at $z<0.2$. This evolution of $ \alpha $ in
fossil groups can be explained by a recent devouring of dwarf galaxies
by a central giant galaxy in fossils.  The $\alpha$ corresponding to
the composite LFs within $ R_{200} $ for both fossil and non-fossil
groups, show no significant evolution with redshift.

We have investigated the evolution of $ \Phi^{*} $ of composite LFs
and find that the $ \Phi^{*} $ of composite LFs of fossil groups
within $ 1/4R_{200} $ and $ 1/2R_{200} $ steeply evolves towards
larger values at redshifts below z$\sim 0.3$. This evolution for the
composite LF of fossils within $ R_{200} $ is slow. Similar to fossil
groups, the $ \Phi^{*} $ of composite LF of control and random groups
 significantly increases with redshift.

 (IV) We find that the number of dwarf galaxies in fossil groups has
 no evolution with redshift since z=1 to present day,while a
 significant evolution is observed for control and random groups, by $
 \sim42\% $ and $ 25\% $, respectively. The constant behavior of a
 number of dwarf galaxies in fossil groups indicates that the changes
 in $ \Phi^{*} $ in these systems are caused by the changes in $ M^{*}
 $ and not in the number of galaxies.

 (V) In contrast to the fossil and to the control groups, we detect that the
 magnitude gap and the composite LF of random groups show no
 considerable evolution with redshift, indicating the importance of
 the magnitude gap as an optical tool in the study of galaxy groups.
  
 Our findings on the time evolution of the Schechter parameters,
 suggest that the central elliptical galaxies in local fossil groups
 have been formed by multiple massive galaxy mergers since
 z$\sim$0.5. Because the timescale for the merging of faint galaxies
 is larger than the timescale for merging of massive galaxies in
 groups, the effect of this merger scenario is more pronounced in the
 bright end of galaxy luminosity functions of fossil groups than in
 the faint end.

\section{Acknowledgements }
    
We thank Saeed Tavasoli, Ali Koohpaee, Amin Farhang, Ehsan Kourkchi,
Anoushiravan Roozrokh, and Mojtaba Raouf for helpful assistance. 
We used the Millennium Simulation databases in this paper
and the web application providing online access to them were
constructed as part of the activities of the German Astrophysical
Virtual Observatory. This work also has been partially supported by
the grant of Finnish Academy of Sciences to the University of Helsinki,
decision number 266918.

\end{document}